\documentclass{article}

\usepackage{epsfig,psfig,epsf,amsmath,amssymb}
\usepackage{colordvi}

\newcommand{\bc}{\begin{center}}
\newcommand{\ec}{\end{center}}   

\newcommand{\bbox}[1]{\pmb{#1}}

\newcommand{\be}{\begin{equation}}
\newcommand{\ee}{\end{equation}}

\def\bea{\begin{eqnarray}}
\def\eea{\end{eqnarray}}
\renewcommand{\Re}{{\rm Re}\,}
\renewcommand{\Im}{{\rm Im}\,}
\newcommand{\bu}{{\bf u}}

\def\bv{{\bf v}}  
\def\bW{{\sf W}} 
\def\bJ{{\sf J}}      
\def\M{{\sf M}}   
\def\para{{\|}}

\def\bxi{\bbox{\xi}} 

\def\bI{{\bf I}}

\renewcommand{\i}{{\rm i}}
\renewcommand{\d}{{\rm d}}
\newcommand{\e}{{\rm e}}

\newcommand{\cc}{{\rm c.c.}}

\begin{document}

\title{\bf Hebbian imprinting {and retrieval} in oscillatory neural networks}

\author{Silvia Scarpetta\\
Department of  Physics  ``E.~R.\ Caianiello''\\
Salerno University, Baronissi (SA), IT\\
and INFM, Sezione di Salerno, (SA), IT 
\and
Zhaoping Li\\
Gatsby Computational Neuroscience Unit\\
UCL, London, UK 
\and
John Hertz\\
Nordita\\ 
DK-2100 Copenhagen \O } 
 
\date{\today}
 
\maketitle

\begin{abstract}

We introduce a model of generalized Hebbian learning and retrieval in 
oscillatory neural networks modeling cortical areas such as hippocampus 
and olfactory cortex.  Recent experiments have shown that synaptic plasticity
depends on spike timing, especially on synapses from excitatory pyramidal 
cells, in hippocampus and in sensory and cerebellar cortex.
Here we study how such plasticity can be used to form memories and 
input representations when the neural dynamics are oscillatory, as is
common in the brain (particularly in the hippocampus and olfactory cortex). 
Learning is assumed to occur in a phase of neural plasticity, in which the 
network is clamped to external teaching signals.  By suitable manipulation 
of the nonlinearity of the neurons or of the oscillation frequencies during 
learning, the model can be made, in a retrieval phase, either to categorize 
new inputs or to map them, in a continuous fashion, onto the space spanned 
by the imprinted patterns.   We identify the first of these possibilities
with the function of olfactory cortex and the second with the observed 
response characteristics of place cells in hippocampus.   We investigate
both kinds of networks analytically and by computer simulations, and we  
link the models with experimental findings, exploring, in particular, 
how the spike timing dependence of the synaptic plasticity constrains the 
computational function of the network and vice versa.

\end{abstract}

\newpage
 
\section{Introduction}

It has long been known that the brain is a dynamical system in which 
non-static activities are common. In particular, oscillatory neural 
activity has been observed and is believed to play significant functional 
roles in, for example, the hippocampus and the olfactory cortex.  
The inputs to these areas can be oscillatory, and the intra-areal 
connections also make these systems prone to intrinsic oscillatory dynamics.
Networks of interacting excitatory and inhibitory neurons (E-I networks) 
are ubiquitous in the brain, and oscillatory activity is not unexpected 
in such networks because of the intrinsically asymmetric character of 
the interactions between excitatory and inhibitory cells.
Recent experimental findings further underscored the importance of dynamics
by showing that long term changes 
in synaptic strengths depend on the relative timing of pre- and 
postsynaptic firing\cite{magee,debanne,biandpoo,markram, CurtisBell}. 
For instance, in neocortical and hippocampal pyramidal 
cells~\cite{magee,debanne,biandpoo, markram, CurtisBell, feldman, debanne2},
the synaptic strength increases (long-term potentiation (LTP)) or decreases 
(long-term depression (LTD)), depending on whether 
the presynaptic spike precedes or follows the postsynaptic one.
This synaptic modification is largest for differences between pre- and 
postsynaptic spike times of order 10 ms.  
Since the scale of this relative timing is comparable 
to the period of neural oscillations, the oscillatory dynamics
should affect the resulting synaptic modifications.
In particular, the relative phases between the oscillating neurons
ought to constrain the synaptic changes that can occur.
These synaptic strengths should, in turn, determine the 
nature of the network dynamics.   
This interplay seems likely to have significant functional 
consequences.

While we have achieved some understanding of the computational 
power of oscillatory networks (see \cite{li-dayan} and references 
therein), they are poorly-understood in comparison with networks
that always  converge to static states, such as feed-forward networks or 
recurrent networks with symmetric connections.
In particular, while we know a lot about appropriate 
learning algorithms for associative memory in symmetrically-connected 
and feedforward networks \cite{HKP}, there is little
previous work on learning, in the
context of the synaptic physiological findings mentioned above,
in asymmetrically connected networks with oscillatory dynamics.
In this paper, we introduce a model for spike-timing dependent learning
in an oscillatory neural network and show how such a network can 
perform associative memory or input representation after learning.

The experimental findings dictate the general form of our model.  It is an 
E-I (excitatory-inhibitory) network, with asymmetric interactions between
excitatory and inhibitory cells, that exhibits input-driven oscillatory 
activity.  We describe the long-term synaptic changes induced by a pair 
of pre- and postsynaptic spikes at times $t_{pre}$  and $t_{post}$
by a function, which we denote $A(\tau)$, of the difference in spike times
$\tau=t_{post} -t_{pre}$. Hence, $A (\tau)$ is positive or negative for
LTP or LTD for a particular $\tau $ value.   
According to the experiments \cite{magee,debanne,biandpoo,markram,CurtisBell,feldman,debanne2},
 $A(\tau)$ varies in different preparations.
For instance, the synapses between hippocampal pyramidal cells have
$A(\tau) >0 $ when $\tau>0$ and  $A(\tau) <0 $ when $\tau<0$
We will consider a general $A(\tau)$ in order to be able to explore 
the consequences of different forms of this function which may be relevant 
to different areas or conditions.  
We study analytically and by simulation how oscillatory activities 
influence synaptic changes and how these changes influence the network 
oscillations and their functions.  In particular, we ask the following 
questions: (1) How can the system
function as an associative memory or as a substrate for a map of an input
space? (2) What constraints do these functions place on the form of 
$A(\tau)$?  (3) What constraints would particular experimental findings 
about $A(\tau)$ impose on the function of networks like this one?

In the next section we present the model E-I network and describe its
dynamics for arbitrary synaptic strengths, making use of a linearized 
analysis.  Section 3 then applies the spike-timing-dependent synaptic 
dynamics to the firing states evoked by oscillatory input.  
We obtain general expressions for the resulting learned synaptic strengths 
and use the linearized theory to study the response properties of the 
network.  We show how the learning rates can be adjusted so that,
 after learning, the network responds strongly (resonantly) to inputs similar 
to those used to drive it during learning and weakly to unlearned inputs.
In addition to this pattern tuning, the system exhibits tuning with respect 
to driving frequency: the response is weakened for driving frequencies 
different from that used during learning.

We show further that, depending on the kind of nonlinearity in the neuronal 
input-output function, the model can perform two qualitatively different 
kinds of computations.  One is associative memory, in which an input to the 
network is categorized by identifying it with the learned
pattern most similar to it.  (Olfactory cortex is believed to operate in
something like this way.)  The other is to make a representation of the
input pattern as a continuous mapping onto the space spanned by the learned
patterns.  For this mode of operation, which we call ``input representation'',
it is not necessary for the network to have learned
explicitly all patterns to which it should respond; it performs
a kind of interpolation between a much smaller number of prototypes. 
(In the hippocampus, we identify these prototypes with place cell fields.)
Section 4 presents the nonlinear analysis of the network for both these cases. 
In Section 5 we examine the consequences of various possible constraints on 
the signs and plasticities of the synapses.  Despite the primitive character 
of the model we use, we believe our findings may have relevance to the dynamics 
of many cortical areas.  
In the final section we discuss our results in the context of other modeling 
and experimental findings, indicating some interesting directions, 
both experimental and theoretical, for future work.

\section{The model}
\label{sec_model}

\begin{figure}[ththththt!]
\setlength{\unitlength}{0.7pt}
\begin{picture}(400, 510) (0, 90)
\put(130, 650) {\bf A:}

\put(130, 420) {\epsfxsize=145pt \epsfbox{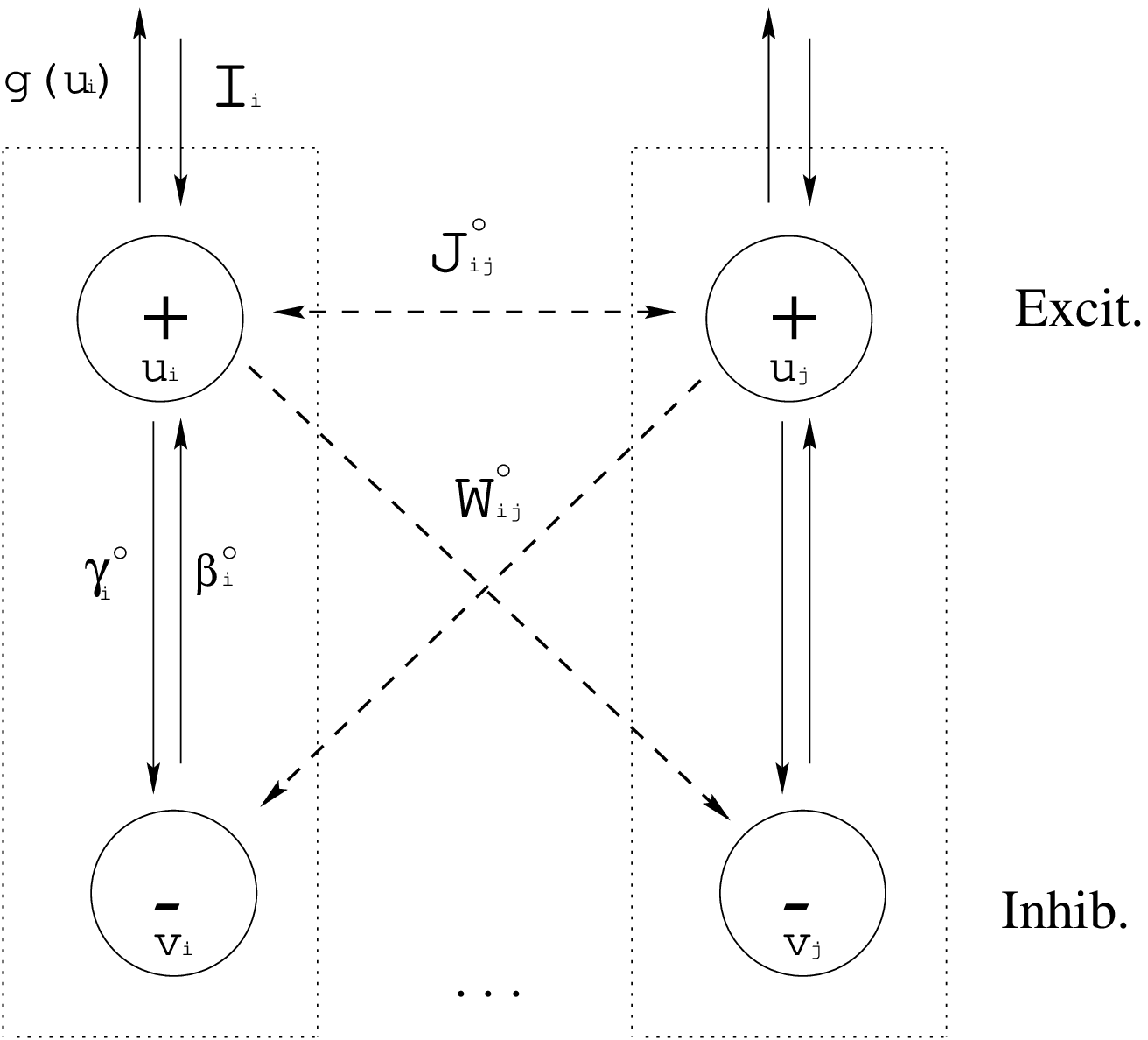}}

\put(5, 378){ {\bf B:} class I  $g_u$} 
\put(185, 378){{\bf C}: class II  $g_u$} 
\put(340, 378){{\bf D}: $g_v$} 
\put(30, 260) {\epsfxsize=50pt \epsfbox{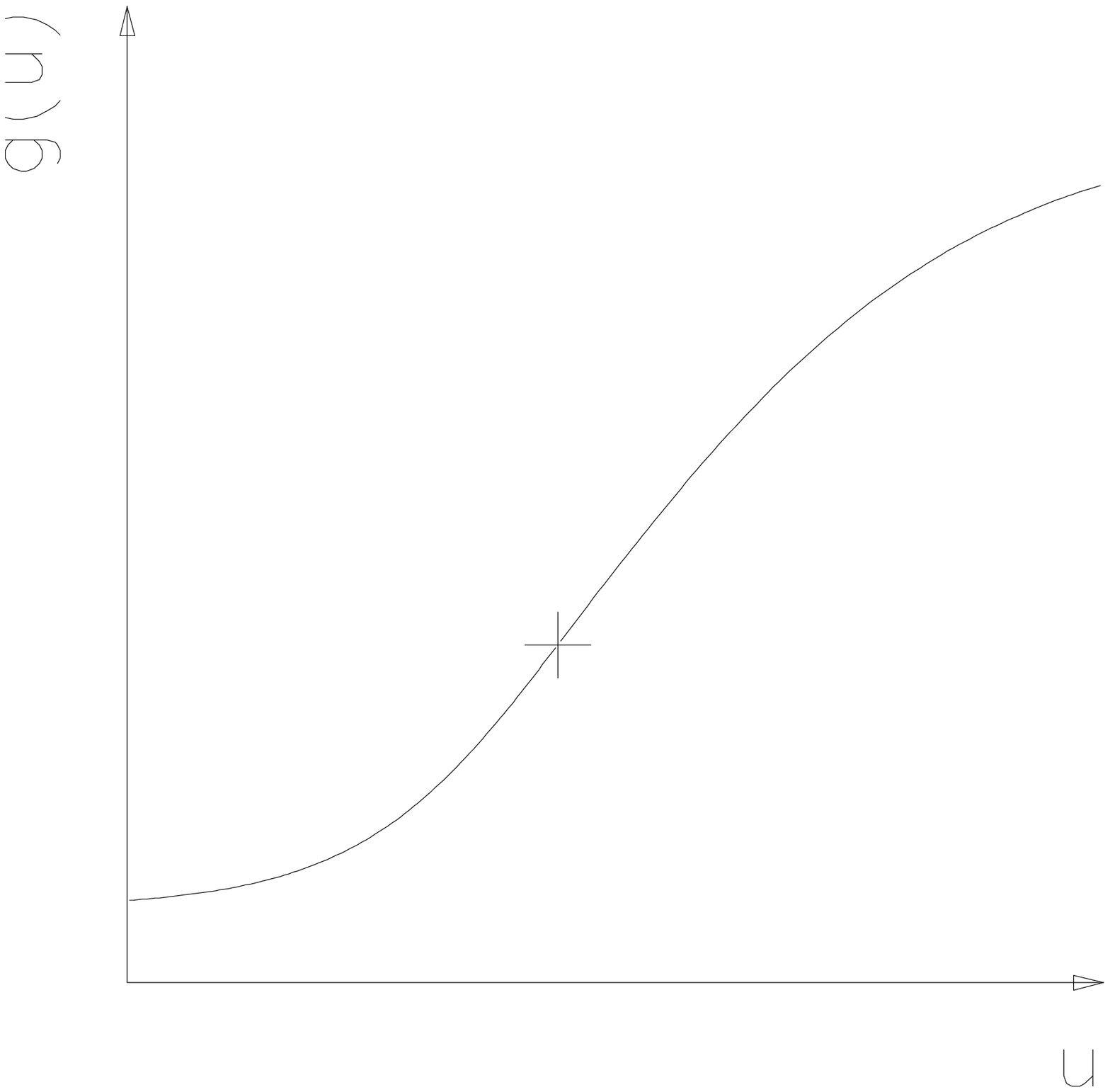}}
\put(205, 260) {\epsfxsize=50pt \epsfbox{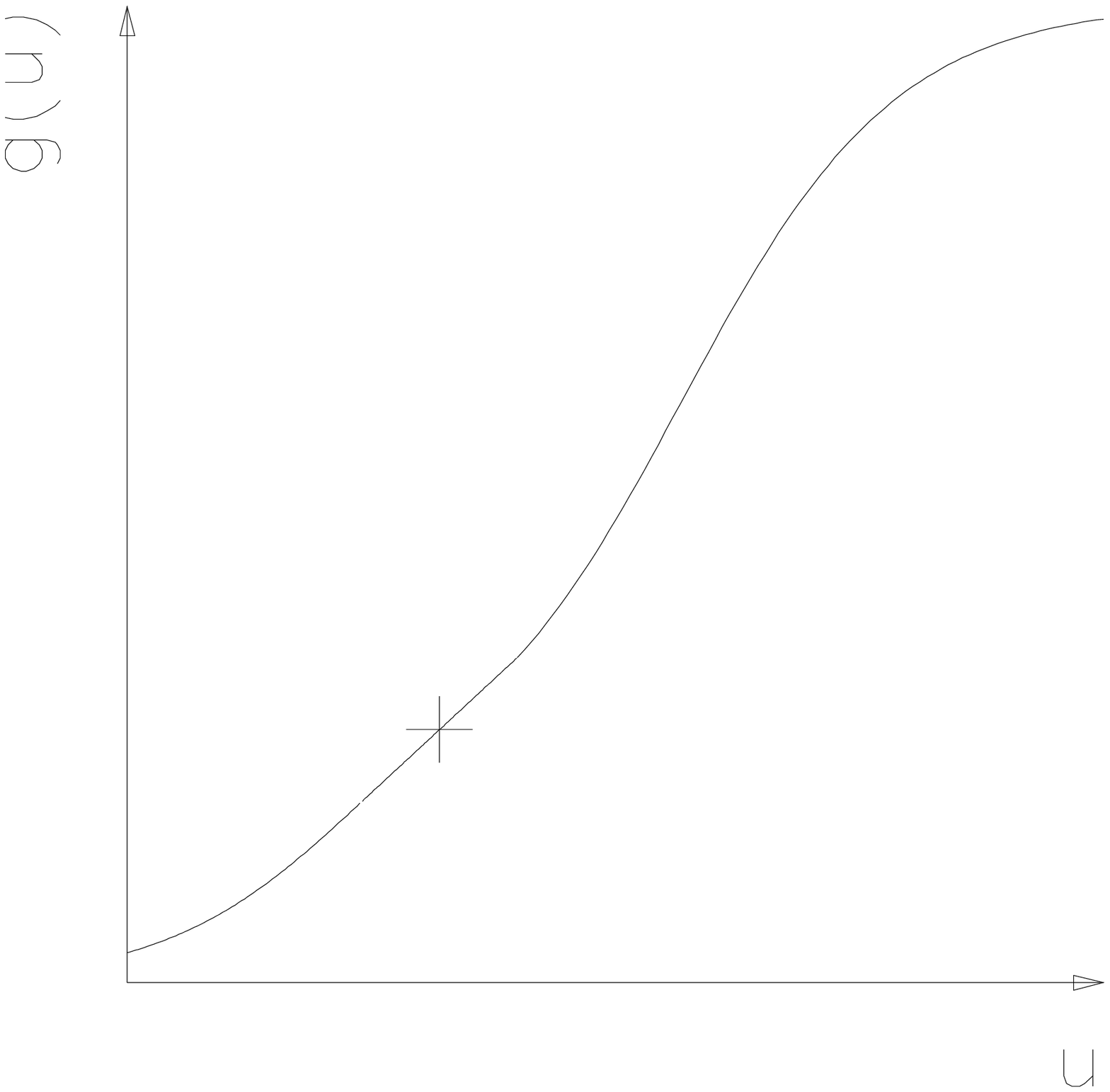}}
\put(355, 260) {\epsfxsize=50pt \epsfbox{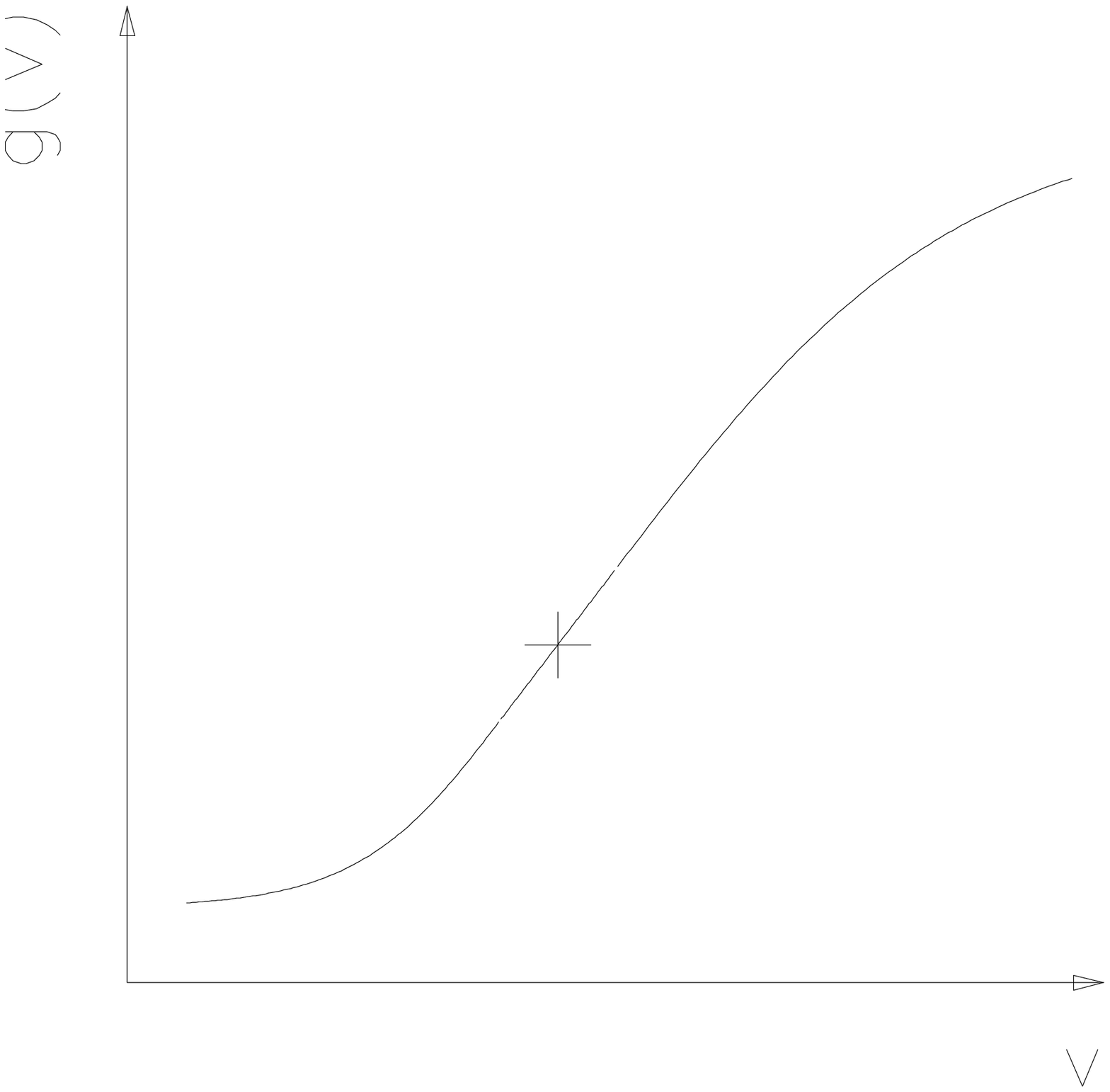}}

\put(5, 220){ {\bf E:} $A(\tau)$} 
\put(190, 220){{\bf F}: $\Re [\tilde A(\omega )]$} 
\put(340, 220){{\bf G}: $\Im [\tilde A(\omega )]$}   
\put(-30, 105) {\epsfxsize=195pt \epsfbox{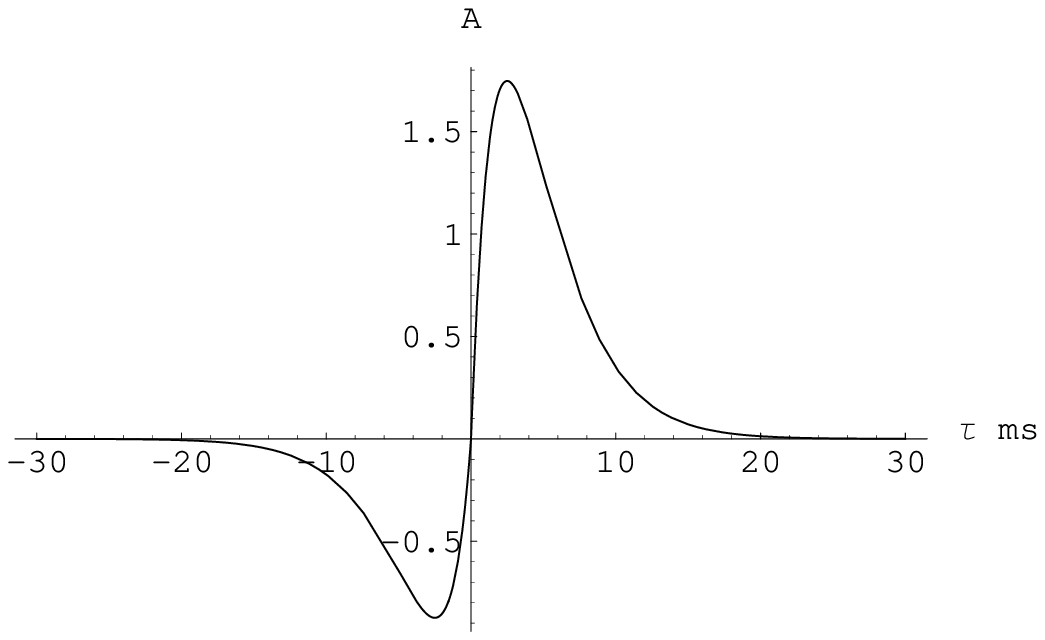}}
\put(180, 70) {\epsfxsize=115pt \epsfbox{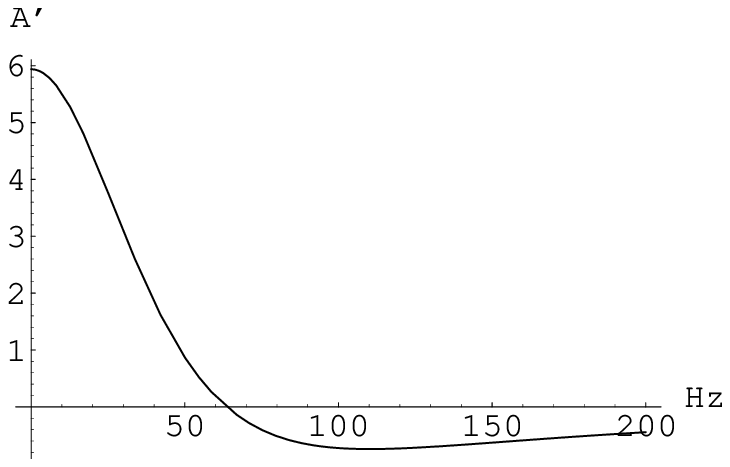}}
\put(335, 70) {\epsfxsize=125pt \epsfbox{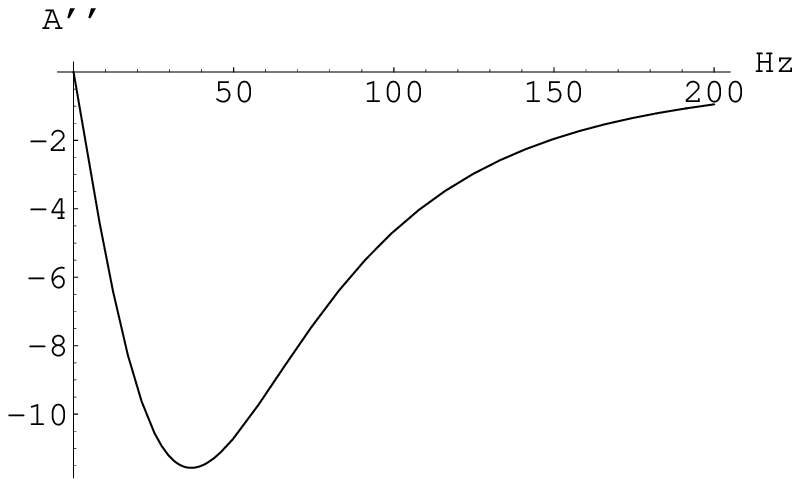}}
\end{picture}
\caption{A: The model elements.  
Input is to the excitatory units $u$,
which also provide the network output.
There are local excitatory-inhibitory connections
(vertical solid lines) and nonlocal connections (indicated by dashed lines) 
between
the excitatory units ($J_{ij}$) and from
excitatory to inhibitory units ($W_{ij}$).
B, C, D: the activation functions for model neurons. Class I and class II 
nonlinearities are shown in B and C respectively.
Crosses mark the equilibrium point $(\bar u, \bar v )$ of the system
(see Sect.\ \ref{linsect}) used in our numerical simulations.
The slopes of all activation functions used in these
calculations are taken to be 1 at the equilibrium point.
E,F,G: an example of kernel shape $A(\tau)$ \protect\cite{kempterPRE} 
and the real 
and imaginary part of its Fourier transform.
}
\label {f_model}
\end{figure}

We base our model on one formulated recently by two of us \cite{LH00} to 
describe olfactory cortex.  For completeness, we summarize its main features 
here.  In the brain regions we model, hippocampus or olfactory cortex,
pyramidal cells make long range connections both to other pyramidal cells 
and to inhibitory interneurons, while inhibitory interneurons generally only 
project locally (Fig.~\ref{f_model}A).  The elementary module of the system 
is an E-I pair consisting of one excitatory and one inhibitory unit, mutually 
connected.  Each such unit represents a local assembly of pyramidal cells or 
local interneurons sharing common, or at least highly correlated, input.
(The number of neurons represented by the excitatory units is in general 
different from the number represented by the inhibitory units.)
Such E-I pairs, without connections between them, form independent damped local
oscillators.  The connections between units in different pairs, which we 
term long range connections, are subject to learning or plasticity in 
this study.  They couple the pairs and determine the normal modes of the  
coupled-oscillator system. 
The input to the system is an oscillating pattern $I_i$ driving the 
excitatory units.
It models the bulb input to the pyramidal cells in the olfactory cortex 
or the input from the enthorinal cortex (perforant path) and
the dentate gyrus (mossy-fiber) to CA3 pyramidal cells. The system outputs 
are from the excitatory cells.

The state variables, modeling the membrane potentials, are
$\bu=\{u_1,\ldots,u_N\}$ and $\bv=\{v_1,\ldots,v_N\}$
respectively for the  excitatory and  inhibitory units.  (We denote vectors 
by bold  font.)  The unit outputs, representing the probabilities of the 
cells firing (or instantaneous firing rates) are given by 
$ g_u(u_1),\ldots,g_u(u_N)$ and $g_v(v_1),\ldots,g_v(v_N)$, where $g_u$ and 
$g_v$ are sigmoidal activation functions that model the neuronal input-output 
relations.  The equations of motion are
\bea
\dot{u}_i&=& -\alpha u_i -\beta_i^{0} g_v(v_i) + \sum_{j} J_{ij}^{0} g_u(u_j) +
 I_i  ,						\label{eqnsu}		\\
\dot{v}_i&=& -\alpha v_i +\gamma_i^{0} g_u(u_i) 
+ \sum_{j \neq i} W_{ij}^{0} g_u(u_j) .		\label{eqnsv} 
\eea
where  $\alpha^{-1}$ is the membrane time constant (for simplicity assumed the same
for excitatory and inhibitory units), $ J_{ij}^{0}$ is the synaptic strength 
from excitatory unit $j$ to excitatory unit $i$, $ W_{ij}^{0}$ is the synaptic 
strength from excitatory unit $j$ to inhibitory unit $i$, $\beta_i^0$ and
$\gamma_i^0$ are the local inhibitory and excitatory connections within the 
E-I pair $i$, and $I_i(t)$ is the net input from other parts of the brain.  
We omit inhibitory connections between pairs here, since the real anatomical 
long-range connections appear to come predominantly from excitatory cells.  
(The parameter $\gamma_i^0$ could be 
identified as $W_{ii}^0$ and the term $\gamma_i^{0} g_u(u_i)$ absorbed into 
the following sum over $j$, but for later convenience, we have written this 
local term explicitly.)  All these parameters are non-negative; the inhibitory 
character of the second term on the right-hand side of Eqn.\ (\ref{eqnsu}) 
is indicated by the minus sign preceding it.

\subsection{linearization}\label{linsect}

The static part $\bar \bI$ of the input determines a fixed point 
$(\bar{\bu},\bar{\bv})$, given by the solution of equations $\dot{\bu}=0, 
\dot{\bv}=0$ with $\bI = \bar \bI$.  
Linearizing the equations 
(\ref{eqnsu}) and (\ref{eqnsv}) around the fixed point leads to  
\bea
\dot{u_i} &=& -\alpha u_i -\beta_i v_i + \sum_j J_{ij} u_j +
 \delta I_i \nonumber \\
\dot{v}_i&=& -\alpha v_i +\gamma_i u_i + \sum_j W_{ij} u_j  
\label{eqnslin}
\eea  
where $u_i$ and $v_i$ are now measured from their fixed point values,
$\delta \bI \equiv \bI - \bar \bI \,$, 
$\beta_i = g'_v(\bar v_i)  \beta_i^{0}$,
$\gamma_i = g'_u(\bar u_i) \gamma_i^{0}$,
$W_{ij}=g'_u(\bar u_j) W_{ij}^{0}$
$J_{ij}=g'_u(\bar u_j) J_{ij}^{0}$.  Henceforth, for simplicity, we assume
$\beta_i = \beta$, $\gamma_i=\gamma$, independent of $i$.

Eliminating the $v_i$ from (\ref{eqnslin}), we have
the  second order differential equations 
\begin{equation}
 \left[(\partial_t +\alpha)^2 +\beta \gamma \right] \bu =
\M  \bu+ (\partial_t +\alpha)\delta \bI
\label{eqlin1}   
\end{equation}
\begin{equation}
\mbox{where} \qquad \M=  (\partial_t +\alpha)  {\sf J}-
\beta {\sf W} \, ,
\label{MMM}
\end{equation}       
or, equivalently,
\begin{equation}
\ddot{\bu} + (2\alpha -{\sf J}) \dot{\bu} + [\alpha^2 - \alpha {\sf J} +
 \beta (\gamma +{\sf W})] \bu 
 = (\partial_t + \alpha)\delta {\bf I}.  			\label{ddu}
\end{equation}
(We use sans serif to denote matrices.)

Given a stable fixed point, an oscillatory drive
$\delta \bI \equiv 
\delta \bI^+ + \delta \bI^-,		 		       
$ where $\delta \bI^+ \propto e^{-\i\omega t }$ and
$\delta \bI^- = \delta \bI^{+*}$, 
will lead eventually to a sustained oscillatory response
$\bu \equiv \bu^+ +\bu^-$ with the same frequency $\omega$, with
$\bu^+\propto \e^{-\i \omega t}$ and $\bu^- = \bu^{+*}$. 
Then from Eqn.\ (\ref{eqlin1}),
\begin{equation}
\left[(-\i \omega + \alpha)^2 +\beta\gamma
\right] u_i^{+}= \sum_j M_{ij} u_j^{+} 
+\left(\alpha -\i \omega \right) \delta I_i^{+},      \label{eqnsM}
\end{equation}
where
\begin{equation}
{\sf M} = (\alpha -\i \omega){\sf J} - \beta {\sf W} .				 
       \label{Mdef}
\end{equation}
is now the ${\sf M}$ in equation (\ref {MMM}) applied to the $e^{-\i\omega t }$ modes.
The terms in the square bracket describe the local E-I pair 
contribution, while ${\sf M}$ gives an effective coupling 
between the oscillating E-I pairs.  
A zero ${\sf M}$ makes ${\bu}$ proportional to $\delta \bI$ with
a constant phase shift, i.e., each individual damping oscillator 
is driven independently by a component of the external drive. 
Learning imprints patterns into $\M$ through the long range connections
$\sf J$ and $\sf W$.  After learning, ${\bu}$ depends on how $\delta \bI^+$ 
decomposes into the eigenvectors of $\M$.  Thus the network can selectively 
amplify  or distort $\delta \bI$ in an imprinted-pattern-specific manner 
and thereby function as an associative memory or input representation.

\subsection{Nonlinearity}

At large response amplitudes, nonlinearity in $g_u$ and $g_v$ significantly 
modifies the response.
We will focus on the nonlinearity in $g_u$ only, since $g_v$ only affects 
the local synaptic input while $g_u$ also affects the long range input
mediated by $\sf J$ and $\sf W$.  We categorize the nonlinearity into
2 general classes in terms of how $g_u$ deviates from linearity near the
fixed point $\bar \bu$:
\begin{equation}      \mbox{class I:} ~~~~~~g_u (u_i) \sim u_i - a u_i^3  
~~~~~~~~~~ \mbox{class II:}
~~~~~~        g_u (u_i) \sim u_i + a u_i^3 - b u_i^5
\label{gclass}
\end{equation}
where  $a, b >0$, 
 and $u_i$  is measured from the fixed point value $\bar u_i$.
Class I and II nonlinearity differ in whether the gain $g'_u$ decreases
or increases (before saturation) as one moves away from the equilibrium point, 
and will lead to qualitatively different behavior, as will be shown.
We will not treat the more general case where $g_u(u)$ is not an odd function 
of $u$.  However, to lowest order a quadratic term just acts to shift the 
equilibrium point, and a quartic one does not 
affect our results qualitatively.

\section{Learning, neural dynamics, and model behavior}
\label{dynamics}

In our treatment, we distinguish a learning mode, in which the 
oscillating patterns are imprinted in the synaptic connections $\sf J$ 
and $\sf W$, from a recall mode, in which connection strengths do not 
change.  Of course, this distinction is somewhat artificial; 
real neural dynamics may not be separated so cleanly into such distinct 
modes.  Nevertheless, cholinergic modulatory effects probably do  
weaken synapses during learning \cite{Hasselmoetc}, 
so there is an experimental basis for the distinction, and it is 
conceptually indispensable.

In what follows we will consider learning of oscillation patterns of
two kinds.  In one, two local oscillators are either in phase with 
each other or $180^\circ$ out of phase, i.e., we can write 
$u_i(t) \propto \xi_i \cos \omega t$, where the $\xi_i$ are real numbers 
(either positive or negative) describing the amplitudes on the different
sites.   In the second
kind of pattern, different local oscillators can have different phases: 
$u_i(t) \propto |\xi_i| \cos (\omega t - \phi_i)$.  We can describe both
cases by writing $u_i(t) = \xi_i \e^{-\i \omega t} + \cc$, taking the 
$\xi_i$ real in the first case and complex ($\xi_i = |\xi_i|\e^{\i \phi_i}$)
in the second.  Thus we will often call the first case ``real patterns'' and
the second ``complex patterns''.

\subsection{learning mode}

Let $C_{ij}$ be the synaptic strength from presynaptic unit $j$  
to postsynaptic unit $i$.  Let $x_j(t)$ and $y_i(t)$ represent the 
corresponding activities relative to some stationary levels at 
which no changes in synaptic strength occur.  Then $C_{ij}$ changes 
during the learning interval $[0,T]$ according to 
\begin{equation} 
\delta C_{ij}(t) = \langle y_i(t)  A(t-t^\prime) x_j(t^\prime) \rangle = 
\frac{\eta}{T} \int_0^{T}\d t \int_{-\infty}^{\infty}\d t^\prime\, 
y_i(t)  A(t-t^\prime) x_j(t^\prime) .  			\label{lr}
\end{equation}
where $\eta$ is the learning rate
and $T$ may be taken  equal to the period of the oscillating input.
The kernel $A(t-t^{\prime})$ is the measure of the strength of synaptic 
change at time delay $\tau=t -t^{\prime}$.    
For example, conventional Hebbian learning, with $A(\tau) \propto \delta 
(\tau )$ (used, e.g., in \cite{LH00}), gives $\delta C_{ij} \propto \int_0^{T}
\d t \, u_i(t)u_j(t)$.  Some experiments \cite{biandpoo,markram} 
suggest $A(\tau)$ to be a nearly antisymmetric function of $\tau$, positive 
(LTP) for $\tau>0$ and negative (LTD) for $\tau<0$ (see Fig.~\ref{f_model}EFG). 
However, for the moment we do not restrict its shape.

Applying the learning rule to our connections $J_{ij}$ and $W_{ij}$, we use 
Eqn.\ (\ref{lr}) with ${\bf x} = {\bf u}$, ${\sf C} = {\sf J}$ or 
$\sf W$, and ${\bf y} = {\bf u}$ or $\bf v$, respectively, giving:
\bea
J_{ij} &=& \frac{1}{NT}
\int_0^T\d t \int_{-\infty}^{\infty}\d t^\prime\,
u_i(t) A_{J}(t-t^\prime) u_j(t^\prime) 			\nonumber \\
W_{ij} &=& \frac{1}{NT}
\int_0^T\d t \int_{-\infty}^{\infty}\d t^\prime\,
v_i(t) A_{W}(t-t^\prime)u_j(t^\prime).			\label{jw}
\eea
We have absorbed the learning rates into the definition of the kernels
$A_{J,W}$ and added the conventional normalizing factor $1/N$ for 
convenience in doing the mean field calculations.   

As mentioned above, cholinergic modulation can affect the strengths of 
long-range connections in the brain; these are apparently almost ineffective 
during learning \cite{Hasselmoetc}.  The neural dynamics is then simplified 
in our model by turning off $\sf J$ and $\sf W$ (and thus $\M$) in the learning 
phase.

Consider the learning of a  single input pattern,    
$\delta \bI = \bxi^{\mu} e^{-\i \omega_\mu t}+ \cc$.  
We calculate separately the responses $u_i^{\pm}$ and $v_i^{\pm}$ to the 
positive- and negative-frequency parts of the input, add them together,
and use the resulting $u_i(t)$ and $v_i(t)$ in Eqns.\ (\ref{jw}) to 
calculate $J_{ij}$ and $W_{ij}$.
For the positive-frequency response we obtain   
\bea
u_i^{+} &=& 
 \frac{( \alpha -\i \omega_{\mu}) \xi_i^{\mu}\e^{-\i  \omega_{\mu} t}
 }{
 ( \alpha -\i \omega_{\mu})^2 + \beta \gamma 
 }
 \equiv \chi_0(\omega_\mu)
\xi_i^{\mu} \e^{-\i \omega_{\mu} t } \label{ui+} \\
v_i^{+} &=& \frac{\gamma }{\alpha -\i \omega_{\mu}} 
\chi_0(\omega_\mu) \xi_i^{\mu} \e^{-\i \omega_{\mu} t }.  
\label{vi+}
\eea 
The quantity $\chi_0(\omega)$ is the output-to-input ratio
for the network with $\sf J$ and $\sf W$ equal to zero. 
The responses $u_i^-$ and $v_i^-$ to the corresponding negative-frequency
driving pattern $\delta \bI^- = \bxi^{\mu *} e^{\i \omega_\mu t}$ are
the complex conjugates of (\ref{ui+}) and (\ref{vi+}), respectively. 

Substituting these responses into Eqns.\ (\ref{jw}) yields connections
\bea
J_{ij}^{\mu}  &=&  \frac{2}{N} 
\Re [\tilde A_{J}({\omega_{\mu}}) \,\xi_i^{\mu} \xi_j^{\mu *} ] 
\nonumber \\
W_{ij}^{\mu}  &=& \frac{2\gamma}{N}   \Re\big[ \frac{
\tilde{A_{W}}({\omega_{\mu}})}{\alpha -\i \omega_\mu} \xi_i^{\mu}
\xi_j^{\mu *} \big] ,
\label{JWlearning} 
\eea
where 
\begin{equation}
\tilde A_{J,W}(\omega) = |\chi_0(\omega)|^2 \int_{-\infty}^{\infty} 
\d \tau\,A_{J,W}(\tau) \e^{-\i \omega \tau}		\label{Atilde}
\end{equation}
can be thought of as an effective learning rate at a frequency $\omega$. 
The factor of the Fourier transform of the learning kernel carries the
information about different efficacies of learning for different 
postsynaptic-presynaptic spike time differences, while the factor 
$|\chi_0(\omega)|^2$ reflects the responsiveness of the uncoupled
local oscillators (${\sf J} = {\sf W} = 0$) in the learning phase.   
Note that $\Im \tilde A_{J,W}(\omega) =0$ if $A_{J,W}(\tau )$ is
symmetric in $\tau$ and that $\Re \tilde A_{J,W}(\omega) =0$ if  
$A_{J,W}(\tau )$ is antisymmetric. 
We will sometimes denote the real and imaginary parts of $\tilde A_{J,W}$  by 
$\tilde A'_{J,W}$ and $\tilde A''_{J,W}$, respectively. 

The resulting effective coupling $\M$ between oscillators after learning, 
under positive-frequency external drive $\delta \bI^+$ of frequency 
$\omega >0$ (in general $\omega \ne \omega_\mu$), is 
\begin{equation}
M_{ij}^{\mu} = \frac{2(\alpha -\i \omega)}{N}  
\Re [\tilde A_{J}({\omega_{\mu}})\xi_i^{\mu} \xi_j^{\mu *}]
- \frac{2 \beta \gamma}{N} \,
\Re \big[ \frac{\tilde A_{W}(\omega_{\mu}) }{\alpha -\i \omega_\mu}
   \xi_i^{\mu}  \xi_j^{\mu *} \big] 
\label{general_M}
\end{equation} 

The dependence of the neural connections ${\sf J}$ and ${\sf W}$ and
the oscillator couplings $\M$ on $\xi_i^{\mu} \xi_j^{\mu *}$ is just a natural 
generalization of the Hebb-Hopfield factor $\xi_i^{\mu} \xi_j^{\mu}$ 
for (real) static patterns.  This becomes particularly clear if we 
consider the special case when there is the following matching condition
between the two kernels:
\begin{equation}
\tilde A_{J}(\omega_\mu)=
\frac{ \beta \gamma }
{ \alpha^2 + \omega_{\mu}^2}  \tilde A_{W}(\omega_\mu), ~~~~~~~ 
\omega = \omega_\mu.
\label{match_case}
\end{equation} 
Then the oscillator coupling simplifies into a familiar outer-product form for 
complex vectors $\bxi$:
\begin{equation}
M_{ij}^\mu = 
-2 \i \omega_\mu  
 \tilde A_{J} (\omega_\mu )\xi_i^{\mu} \xi_j^{\mu *}/N, 
						\label{Msimple}
\end{equation}

To construct the corresponding matrices for multiple patterns (which we will
always take to be random and independent)
we simply sum (\ref{JWlearning}) over input patterns, labeled by 
the index $\mu$, as for the Hopfield model.   We restrict attention to the
case where the number $P$ of stored patterns is negligible in comparison with
$N$, the size of the network (though it may be $\gg 1$).
So far, all our results apply for both real and complex patterns.

\subsection{Recall mode}
\label{sec_recall}

After learning, the connections are fixed and the response $\bu^+$ to an input
$\delta \bI^+ \propto \e^{-\i \omega t}$ is  described by Eqn.\ (\ref{eqnsM}).
To solve it, we need to know how the $\M$ matrix acts on input vectors.
We consider uncorrelated patterns all learned at the same frequency 
($\omega_\mu$ independent of $\mu$).  Then it is easy to see that $\M$ is a 
projector onto the space spanned by the imprinted patterns.  It has $P$ 
eigenvectors (the imprinted patterns) with the same nonzero eigenvalue and 
$N-P$ with eigenvalue zero.  These are standard properties of outer-product 
constructions for orthogonal vectors; we can treat our $\bxi^\mu$ as
effectively orthogonal here because we are taking the components $\xi_i^\mu$ 
to be independent and $N \gg P$ \cite{AGS}.  

The nonvanishing eigenvalue of $\M$, which we denote $\Pi(\omega, \omega_\mu)$,
is simply computed as
\begin{equation}
\Pi(\omega;\omega_\mu) =  
(\alpha -\i \omega)   \tilde A_{\rm J}(\omega_\mu)
- \frac{\beta \gamma  \tilde A_{\rm W}(\omega_\mu)}
{ \alpha - \i \omega_\mu }   			\label{Mu+complex},
\end{equation} 
for complex patterns, and
\begin{equation}
\Pi(\omega;\omega_\mu) 
= 2 (\alpha -\i \omega) \Re \tilde A_{\rm J}(\omega_\mu)  
- 2 \beta \gamma  \Re\big[ \frac{
\tilde A_{\rm W}(\omega_\mu)}{\alpha - \i \omega_\mu} \big]   \label{Mu+real}
\end{equation}  
for real patterns.

Thus, from Eqn.\ (\ref{eqnsM}), the response $\bu^+$ to an input 
$\delta \bI^+$ in the imprinted-pattern subspace is
\begin{equation}
\bu^+ = \chi (\omega; \omega_{\mu}) \delta \bI^+,         \label{linresp}
\end{equation}
with the linear response coefficient or susceptibility
\begin{equation}
\chi (\omega; \omega_{\mu}) = \frac{\alpha - \i \omega}{ \alpha^2 + \beta
\gamma - \omega^2 - 2\i \omega \alpha - \Pi(\omega;\omega_\mu)}.
\label{chi}
\end{equation}

To achieve a resonant response to an input at the imprinting frequency 
($\omega = \omega_\mu$), the learning kernels should be adjusted so 
that both the real and imaginary parts of the denominator in 
$\chi (\omega _\mu ; \omega _\mu)$ are close to zero,
 i.e.,
\begin{eqnarray}
\epsilon &\equiv & 
\alpha^2 + \beta\gamma -\omega^2_\mu - \Re \Pi(\omega _\mu;\omega_\mu) 
                    \rightarrow       0,      \label{res_c} \\
\Delta &\equiv  &2\omega_\mu  \alpha
 + \Im \Pi (\omega_\mu;\omega_\mu)  \rightarrow 0.
 \label{big_c}
\end{eqnarray}

For real patterns, $\Im \Pi (\omega_\mu; \omega_\mu) = -2\omega_\mu
\tilde A_J'$, so $\Delta = 2\omega_\mu(\alpha - \tilde A_J')$.  Thus
small $\Delta$ requires a positive $\tilde A'_J>0$, i.e., stronger 
positive-$\tau$ LTP than negative-$\tau$ LTD for excitatory-excitatory
couplings (again, provided the typical values of $\tau$ for which 
$A_{J}(\tau)$ is sizeable are small compared to the oscillation period).
>From Eqn.~(\ref{big_c}),
\begin{equation}
\epsilon = \alpha^2 + \beta\gamma -\omega^2_\mu 
- 2 \Re \left[ \alpha \tilde A_J(\omega_\mu)
- \frac{\beta\gamma \tilde A_W(\omega_\mu)}
{\alpha - \i \omega_\mu} \right]			\label{epsreal}
\end{equation}
Thus, for a given $\omega_\mu$, the resonance condition enforces a 
constraint on a linear combination of $\tilde A_J'$, $\tilde A_W'$ and  
$\tilde A_W''$.  However, we note that $\tilde A_J''$ does not 
appear anywhere; it is simply irrelevant to learning real patterns.

For complex patterns,
\begin{eqnarray}
\epsilon & = &
\alpha^2 + \beta\gamma -\omega^2_\mu - 
 (\alpha \tilde A '_J + \omega_\mu \tilde A''_J) 
	+ {\frac{\beta\gamma  } {\alpha ^2 +\omega_\mu ^2 }} 
	(\alpha \tilde A'_W - \omega_\mu \tilde A''_W )  \label{epcomplex}
\\
\Delta &=& 2\omega_\mu \alpha + 
 (\alpha \tilde A ''_J - \omega_\mu \tilde A'_J )
        - {\frac{\beta\gamma  } {\alpha ^2 +\omega_\mu ^2 }} 
        (\alpha \tilde A''_W + \omega_\mu \tilde A'_W )	\label{delcomplex}
\end{eqnarray}
(We have temporarily suppressed the $\omega_\mu$-dependence of 
$\tilde A_{J,W}'$ and $\tilde A_{J,W}''$ to save space.)  One can get
some insight here by considering the time-shifted learning kernels 
$A_J(\tau + \theta_\mu/\omega_\mu)$ and $A_W(\tau - \theta_\mu/\omega_\mu)$,
where $\theta_\mu = \tan^{-1}(\alpha/\omega_\mu)$.  (For $\alpha \sim 
\omega_\mu \approx 40$ Hz, these shifts are around 3 ms.)  In terms of
the associated frequency-domain quantities,
\begin{equation}
\tilde B_{J,W}(\omega) = |\chi_0(\omega)|^2 \int_{-\infty}^{\infty} 
\d \tau\,A_{J,W}(\tau \pm \theta_\mu/\omega_\mu) 
\e^{-\i \omega \tau} = \tilde A_{J,W}(\omega) \e^{\pm \i \theta_\mu},
						\label{Btilde}
\end{equation}
we can write Eqns.~(\ref{epcomplex}) and (\ref{delcomplex}) as
\begin{eqnarray}
\epsilon & =&
\alpha^2 + \beta\gamma -\omega^2_\mu - \sqrt{\alpha^2 + \omega_\mu^2}
\tilde B_J''(\omega_\mu) 
-\frac{\beta \gamma}{\sqrt{\alpha^2 + \omega_\mu^2}}
\tilde B_W''(\omega_\mu)				\label{epcomplexB}	
\\
\Delta &=& 2\omega_\mu \alpha - \sqrt{\alpha^2 + \omega_\mu^2} 
\tilde B_J'(\omega_\mu)
- \frac{\beta \gamma}{\sqrt{\alpha^2 + \omega_\mu^2}}
\tilde B_W'(\omega_\mu).				\label{delcomplexB}
\end{eqnarray}	
Thus, the imaginary parts of the frequency-domain kernels 
$\tilde B_{J,W}(\omega_\mu)$ shift the resonant frequency and the real parts
control the damping.  In particular, one needs at least one of 
$\tilde B'_{J,W}$ to be positive to achieve good frequency tuning.
Negative (positive) imaginary parts $\tilde B''_{J,W}$  
increase (decrease) the resonant frequency.

When the learning window widths in the kernels $A_{J,W}(\tau)$ are much
smaller than the oscillation period, the shifts by $\pm \theta_\mu/\omega_\mu$
do not affect the real parts of $B_{J,W}$ strongly.  However, for window 
shapes (see Fig.~1E) that change rapidly from negative to positive around 
$\tau=0$, the imaginary parts can be strongly suppressed, even for fairly
small shifts.

The explicit form of the resonant response can be seen by expanding
the denominator of $\chi(\omega;\omega_\mu)$ around $\omega=\omega_\mu$:
\begin{equation}
\chi (\omega; \omega_{\mu}) = \frac{\alpha - \i \omega_\mu}
{ \epsilon - i\Delta -Z(\omega_\mu)(\omega -\omega_\mu)},   \label{chi2} 
\end{equation}
where
\begin{equation}
Z(\omega_\mu) = 2\omega_\mu + 2\i \alpha + \left.
\frac{\partial \Pi(\omega;\omega_\mu)}{\partial \omega}\right|_{\omega=
\omega_\mu}.					\label{Zfactor}
\end{equation}
Thus $\chi$ has a pole at 
\begin{equation}
\omega =\omega_\mu + \frac{\epsilon-\i\Delta}{Z}
= \omega_\mu +\frac{\epsilon Z' - \Delta Z'' - \i(\Delta Z' +\epsilon Z'')}
{|Z|^2}, 
\label{poleat}
\end{equation}
and, as the driving frequency $\omega$ in the recall phase is varied, the
system exhibits a resonant tuning, with a peak near $\omega_\mu$ and a 
linewidth equal to  $(\Delta Z' +\epsilon Z'')/|Z|^2$.  

One has to check that the desired learning rates and kernels do not violate 
the condition that the response function (\ref{chi}) be causal, i.e., 
small perturbations decay in time.  Analytically, the requirement is that 
all singularities of $\chi(\omega, \omega_\mu)$ must lie in the lower half of 
the complex $\omega$ plane.  Thus, in Eqn.\ (\ref{poleat}), we need 
$\Delta Z' +\epsilon Z''$ to be positive.  

For real patterns, the analysis is fairly simple.  From Eqns.\ (\ref{Mu+real})
(\ref{big_c}) and (\ref{Zfactor}), we obtain  $Z = 2\omega_\mu +2\i (\alpha
-\tilde A_J')$ and $\Delta = 2 \omega_\mu (\alpha -\tilde A_J')$.  Thus, for
$\Delta \rightarrow 0$, $Z \rightarrow 2\omega_\mu$, and the stability
condition is simply that $\Delta$ be positive, i.e., $\tilde A_J'(\omega_\mu)
< \alpha$.   

For complex patterns, we get $Z = 2\omega_\mu +\tilde A_J'' 
+\i(2\alpha - \tilde A_J')$.  Requiring $\Delta Z' +\epsilon Z''$ to be 
positive then imposes constraints on the signs and relative magnitudes
of $\epsilon$ and $\Delta$, depending on $Z'$ and $Z''$.  We omit the details.

Notice that, for both the real and complex cases, the stability analysis 
does not depend on the $\sf W$-learning kernel $A_W$ at all (except insofar 
as it affects $\epsilon$ and $\Delta$).  This is because $Z$ involves the 
derivative $\partial \Pi(\omega,\omega_\mu)/\partial \omega$, and in both 
cases the only $\omega$-dependence of $\Pi$ is in the factors 
$(\alpha - \i \omega)$ in the first terms of Eqns.~(\ref{Mu+complex}) and
(\ref{Mu+real}), which do not involve $\tilde A_W$.

\begin{figure}[thththth!]
\bc
A 

\vspace{0.2cm}

\begin{tabular}{cccc}
$\omega=\omega_{\mu} $= 33 Hz &
$\omega= $ 38 Hz &
$\omega= $ 41 Hz &
$\omega= $ 49 Hz  
\\
 \epsfysize=6.2cm
\epsfbox{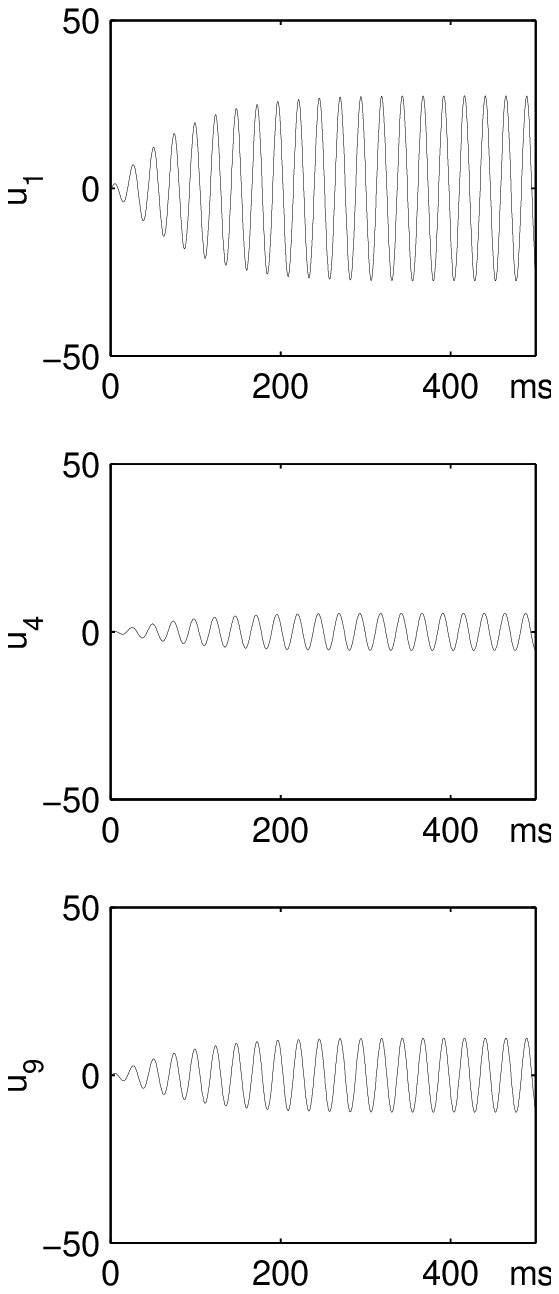} &
\epsfysize=6.2cm
\epsfbox{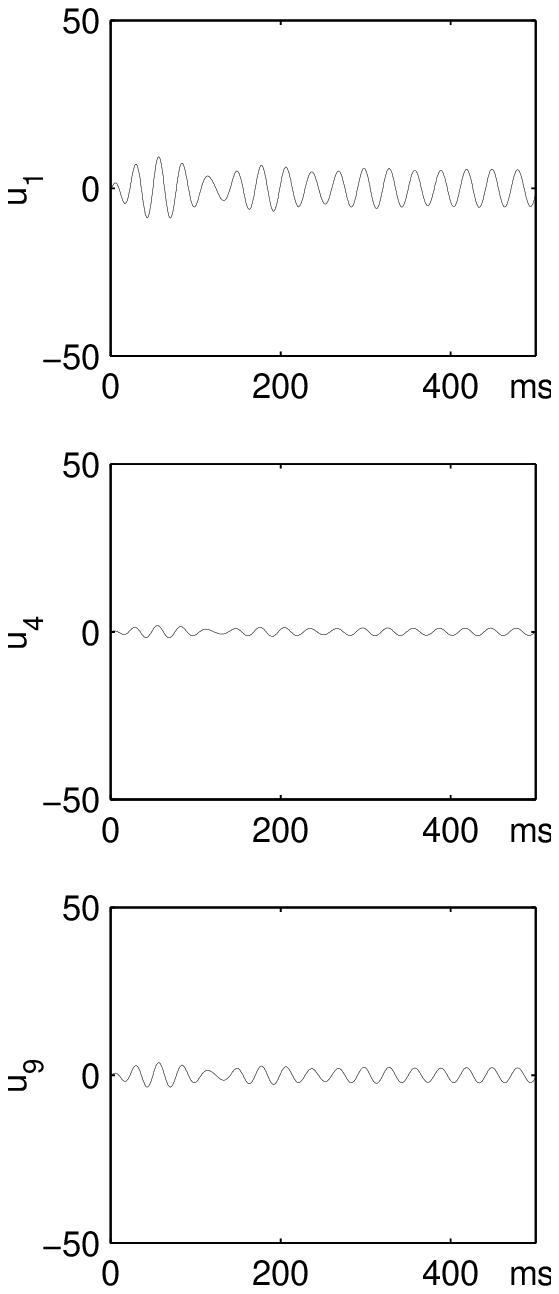} &
\epsfysize=6.2cm
\epsfbox{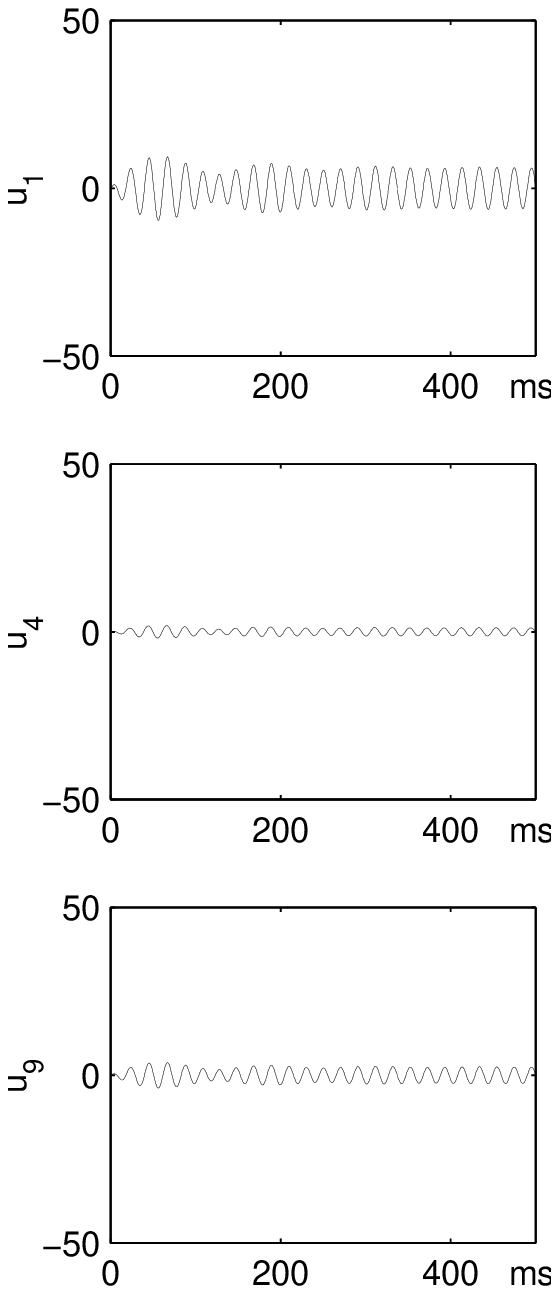} &
\epsfysize=6.2cm
\epsfbox{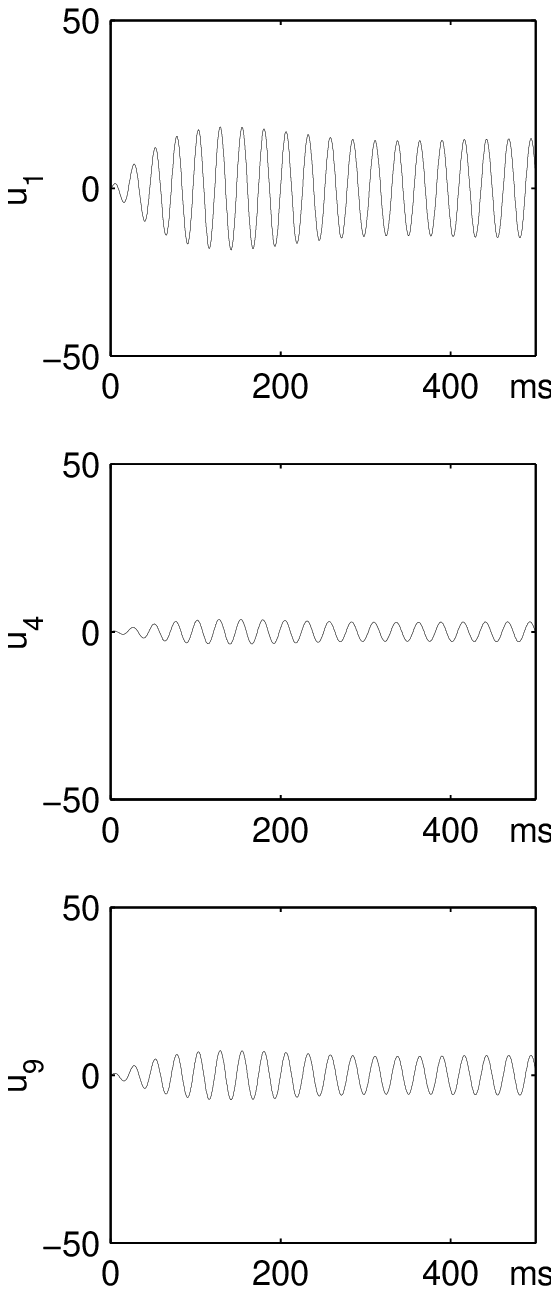} 
\end{tabular}                
\ec
\bc
\begin{tabular}{cc}
B.1 \hspace{1.8cm} & B.2\\
\epsfysize=4.0cm
\epsfbox{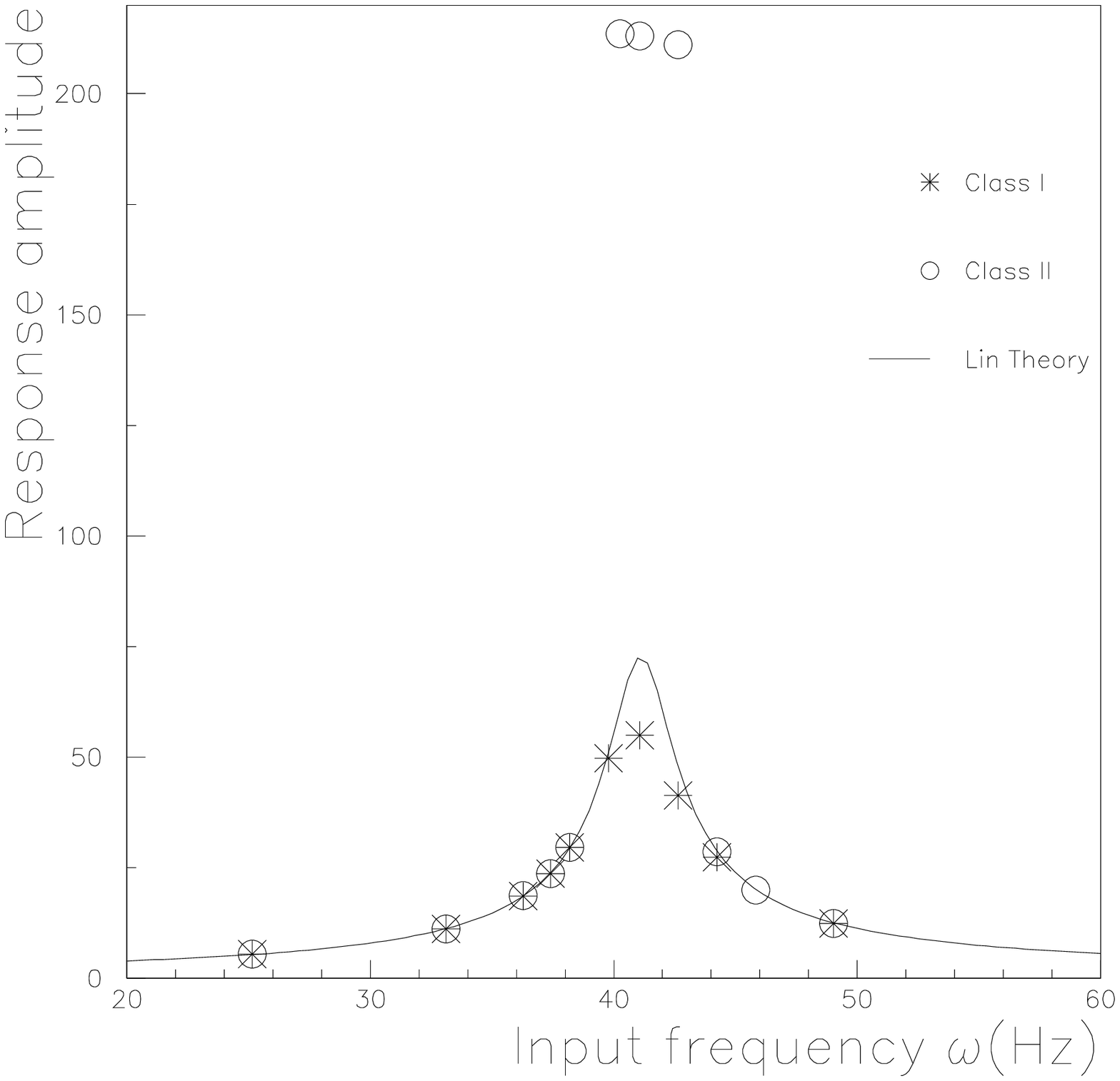} \hspace{2.0cm} &
\epsfysize=4.0cm
\epsfbox{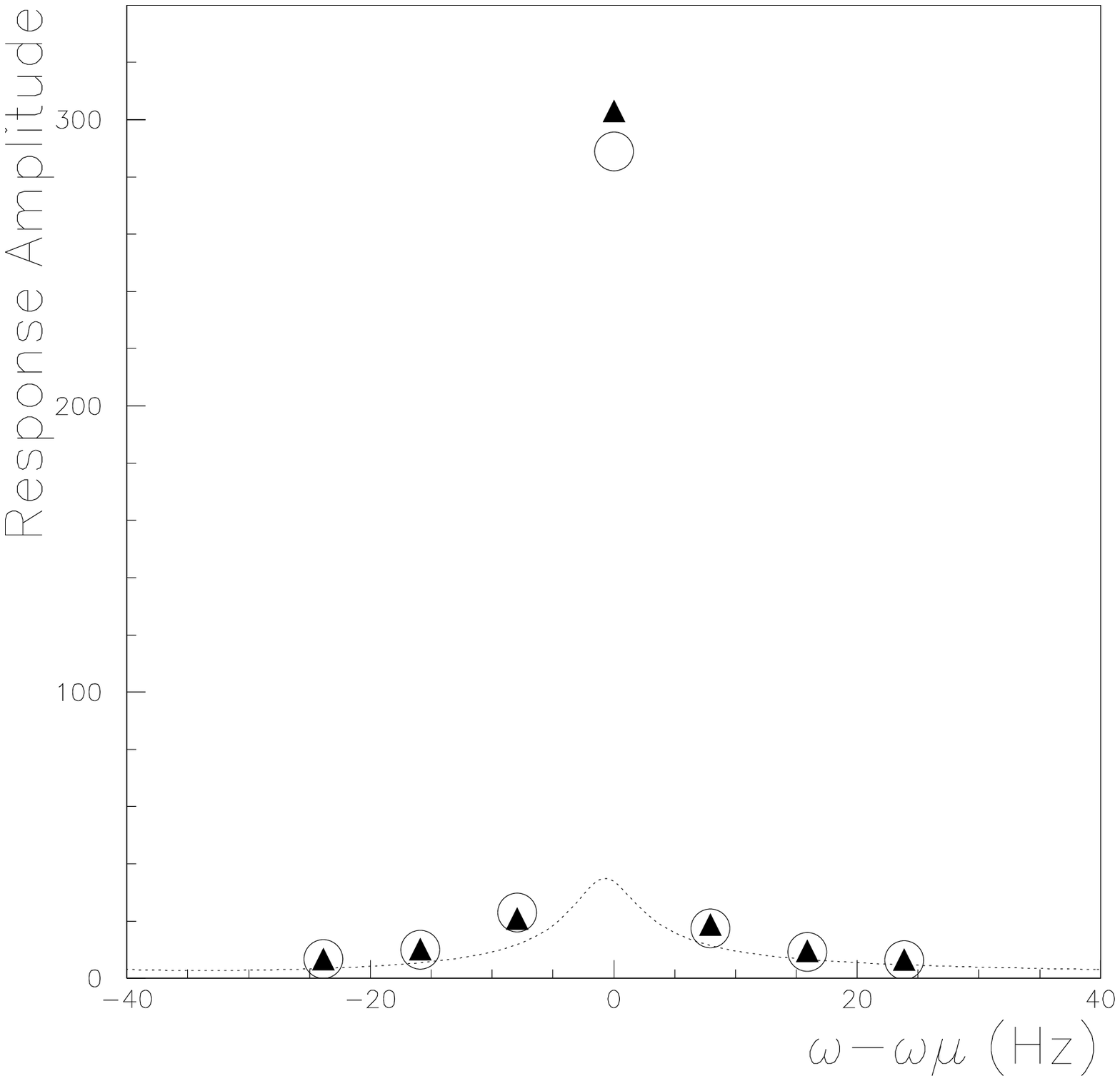}
\\
\end{tabular}                
\vspace{0.1cm}

\caption{
Frequency tuning, shown as response to
$\delta \bI^+ = \bxi_\mu \e^{-\i \omega t}$ after imprinting 
$\bxi_\mu \e^{-\i \omega_\mu t}$ 
with $\omega _\mu =$ 41 Hz.  
The network has 10 excitatory and 10 inhibitory units.
In all figures of this paper
except where explicitly stated, learning kernels are matched 
so that  $\tilde A_{\rm J}
=  \frac{\beta \gamma}{\alpha^2 + \omega_\mu^2}
  \tilde A_{\rm W}  = 0.5 - \i \,  0.028$, and 
$\omega_\mu \sim $ 41 Hz.
A: temporal activities of 3 of the 10 excitatory cells.
$g_u$ and $g_v$ are as in  Fig.\ \ref{f_model}BD.
\quad 
B: frequency tuning curve. Response amplitude  $|\langle \bu^+ |\bxi^\mu
\e^{-\i\omega t}\rangle |$  
(simply $|\chi(\omega;\omega_{{\mu}})|$ in the linearized theory)  to input
$ \delta \bI^+ = \bxi^\mu   \e^{-\i\omega t}$.  B.1: using matched kernel,
from linearized theory (solid line) and from class I (stars) and II
(circles) nonlinearities.
B.2: Opposite-plasticities case  
($\tilde A_{\rm J}= - \tilde A_{\rm W}  = 0.99
( \alpha -\i  \omega_\mu) $), complex patterns.
Dashed line, circles, and triangles are
results from the linearized theory and the class II nonlinear model
with $\omega_\mu= 41 Hz$ and $\omega_\mu= 68 Hz$ respectively. 
}
\label{freqtuning}
\ec
\end{figure}        

Fig.\ \ref{freqtuning} shows examples of the frequency tuning described by 
Eqn.\ (\ref{chi2}), as obtained from simulations of small networks, including 
the nonlinearities described in Sect.\ 2.2.    
Nonlinearity makes the response deviate from the linear prediction when 
the amplitude is larger, as happens near the resonance frequency.  In 
particular, class I and class II nonlinearities lead to reduced and enhanced 
responses, respectively relative to the linear prediction, as will be 
analyzed in detail later.

\subsubsection{Examples of plasticity}

We use examples to illustrate the constraints that resonance and stability 
conditions place on the shape of the kernels for complex patterns.  

\subsubsection*{$\sf J$ only}

If the patterns are imprinted only in the excitatory-excitatory connections,
we have only the first terms on the right-hand sides of Eqns.\ 
(\ref{Mu+complex}) and (\ref{Mu+real}).  

For real patterns  
$\Delta = 2\omega_\mu (\alpha -\tilde  A_J')$ (no different from the general 
case), so, using $\Delta \rightarrow 0$ in Eqn.\ (\ref{epsreal}) yields 
$\epsilon = \beta\gamma -\alpha^2-\omega^2_\mu $.  This then  (from 
$\epsilon = 0$) fixes the imprinting frequency $\omega_\mu = 
\sqrt{\beta \gamma - \alpha^2}$. 

For complex patterns, we have  $\Delta = 2\omega_\mu \alpha - 
\sqrt{\alpha^2 + \omega_\mu^2} \tilde B_J' $, i.e., similar to the 
real-pattern case but with a shifted learning window and a different 
effective strength.  The resonant
frequency shifts down or up according to the sign of $\tilde B_J''$.  
 
\subsubsection*{Same-sign plasticities in $\bJ$ and $\bW$}
\label{sec_match}

Let us consider the case where the kernels are related by the matching
condition (\ref{match_case}).   While the exact match is clearly a special 
case, the simplification it yields in the algebra permits some insight which
can be expected to carry over qualitatively to other cases where the two
kernels have similar shapes and comparable magnitudes.   Here we find, from
Eqns.\ (\ref{Mu+complex}) and (\ref{Mu+real}),
\begin{equation}
\Pi(\omega_\mu; \omega_\mu) =  -2\i \omega_\mu \tilde A_J(\omega_\mu ), 
\label{Pimatched}
\end{equation}
for both real and complex patterns.
Applying the resonance
 condition equations (\ref {res_c}, \ref {big_c}), we have
\begin{eqnarray}
\tilde A_J''(\omega_\mu)&=&
 \frac{-\omega_\mu^2 +\alpha^2 +\beta \gamma - \epsilon}{2
\omega_\mu}
\label{constr_2} \\
\tilde A_J'(\omega_\mu) &=&  \alpha -\Delta /2\omega_\mu
\label{constr_3}
\end{eqnarray}
Thus $\tilde A_J'(\omega_\mu)$ reduces the effective damping
from $\alpha$ to  $\Delta /2\omega _\mu \ll \alpha $, and this
requires $\tilde A'(\omega_\mu) \approx \alpha >0$.
When the width of the learning kernel $A_J(\tau)$ is much smaller than the 
oscillation period,  $\tilde A_J'(\omega_\mu) \approx \int A_J(\tau) \d \tau $, 
thus a positive $\tilde A_J'(\omega_\mu)$ requires that LTP dominate LTD in 
total strength.  

We observe that a negative $\tilde A_J''(\omega_\mu)$, i.e., an 
$A_J(\tau)$ like that in Fig.\ \ref{f_model}EFG,
forces $\omega_\mu$ to be greater than $\sqrt{\alpha^2+\beta\gamma}$ and 
thus greater than the intrinsic E-I pair frequency $\sqrt{\beta \gamma}$
(a shift in the opposite direction from that in the $\sf J$-only, real-pattern
case).

In general, when the width of $A_J(\tau)$ is not small, 
the resonance frequency has to be determined from equations 
(\ref {constr_2}) and (\ref {constr_3}) by $\tilde A_J''(\omega_\mu ) 
/\tilde A_J'(\omega_\mu ) \approx
(-\omega_\mu^2 +\alpha^2 +\beta \gamma )/2
\alpha \omega_\mu$.

\subsubsection*{Opposite-sign plasticities in $\bJ$ and $\bW$}
\label{sec_opposite}

We turn now to the case where $A_J(\tau)$ and $A_W(\tau)$ have opposite signs 
(for all $\tau$).   
Again, we turn to a particular matching of the magnitudes of the two kernels
to find a simple case that can give some general qualitative insight.
We use our old matching condition, Eqn.\ (\ref{match_case}), but with a 
minus sign.  For complex patterns we now find
$\Pi (\omega_\mu,\omega _\mu) = 2\alpha \tilde A_J(\omega_\mu)$,
and, applying the resonance condition equations (\ref {res_c}, \ref {big_c}), 
\begin{eqnarray}
\tilde A_J'(\omega_\mu)&=&
 \frac{-\omega_\mu^2 +\alpha^2 +\beta \gamma - \epsilon}{2
\alpha}
 \\
 -\tilde A_J''(\omega_\mu) &=&  \omega_\mu  -\Delta /2\alpha 
\label{kernela1}
\end{eqnarray}
Comparing these with equations (\ref {constr_2}) and (\ref {constr_3}) and 
the accompanying analysis, we see that the roles of the real and imaginary
parts of $\tilde A_J$ have been reversed:  Now it is the imaginary part 
that is constrained to be near a fixed value ($-\omega_\mu$) by
the $\Delta \rightarrow 0$ condition, and the real part that enters in
the $\epsilon$ equation.
We note that we need $\tilde A_J''(\omega_\mu) <0$ (i.e. like the case
shown in Fig.\ \ref{f_model}E), in order to obtain a small $\Delta$, and 
that the sign of $\tilde A' (\omega _\mu )$ determines whether the
the resonance frequency is larger or smaller than  
$\sqrt{\alpha^2+\beta\gamma}$.

Another interesting special case for complex patterns 
is when $A_{W}=-A_{J}$, with the particular choice
\begin{equation}
\tilde A_J (\omega_\mu)  = \alpha - \i \omega_\mu  
\label{prescri}
\end{equation}
This leads to the remarkably simple result
\begin{equation}
\chi (\omega; \omega_{\mu})  = \frac{\i}{ \omega - \omega_\mu }. 
\label {chi_J-W}
\end{equation}
That is, the choice (\ref{prescri}) satisfies both constraints, $\epsilon,
\Delta \rightarrow 0$ and, in addition, puts the resonance right at the 
original driving frequency.  Fig.~\ref{freqtuning}{.B.2} shows a frequency 
tuning curve obtained with $\tilde A_J (\omega_\mu)= 0.99 (\alpha - 
\i \omega_\mu)$.

To understand the  prescription  $\tilde A_J(\omega_\mu)
= \alpha - \i \omega_\mu$,
consider an oscillation period much greater than the                                  
temporal width of the learning kernel.
Then $\tilde A_J'(\omega_\mu) \approx \int A(\tau) d\tau$ and
 $\tilde A''(\omega_\mu) \approx -\int A_J(\tau) \omega_\mu \tau d\tau  
\propto -\omega_\mu$.  Thus the prescription just requires 
$\int A_J(\tau) d\tau \approx \alpha >0 $ (LTP dominates LTD in total strength)
and $\int A_J(\tau) \tau d\tau \approx 1 > 0$ (LTP when postsynaptic spikes 
follow presynaptic ones, and LTD for the opposite order). This means   
that $A_J(\tau)$ should look like Fig.\ \ref{f_model}E and $A_W(\tau)$ 
like its negative.

We remark that for real patterns this choice of kernels does not produce 
resonant oscillations; in fact, it leads to instability.

\subsubsection{Pattern selectivity}

\begin{figure}[tthththth!]
\bc
A

\vspace{0.3cm }

\begin{tabular}{cccc}
${\it a}: \xi=\xi^\mu$&
${\it b}: |\xi\cdot\xi^\mu|=0.79$&
${\it c}: |\xi\cdot \xi^\mu|=0.36$ &
 \\
\epsfysize=6.6cm
\epsfbox{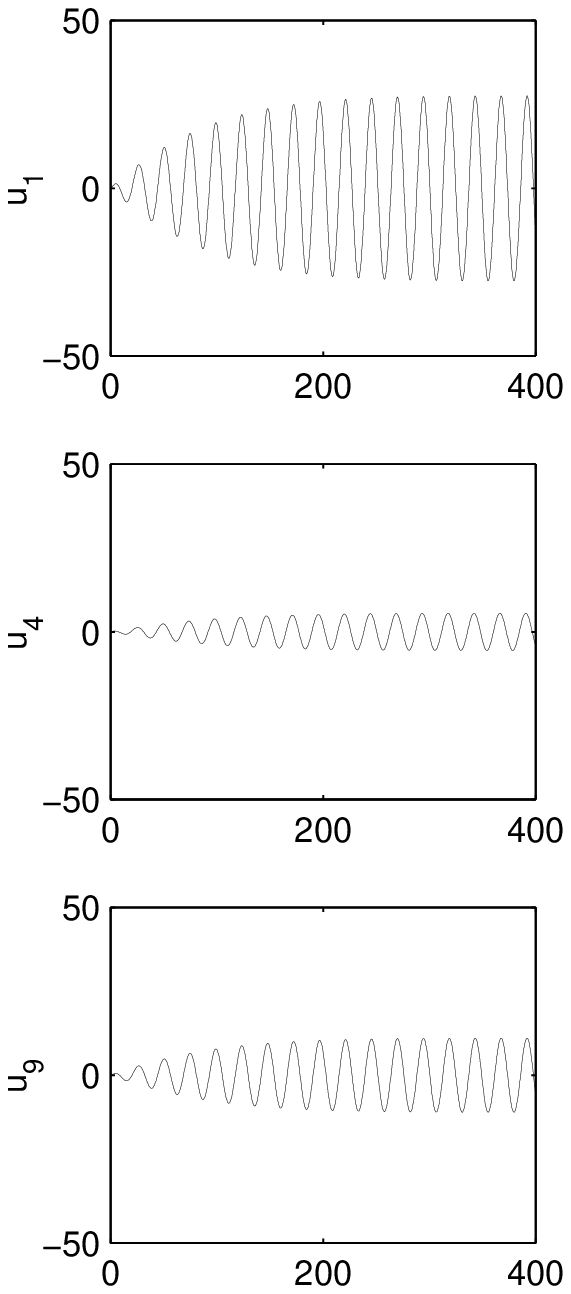}&
\epsfysize=6.6cm
\epsfbox{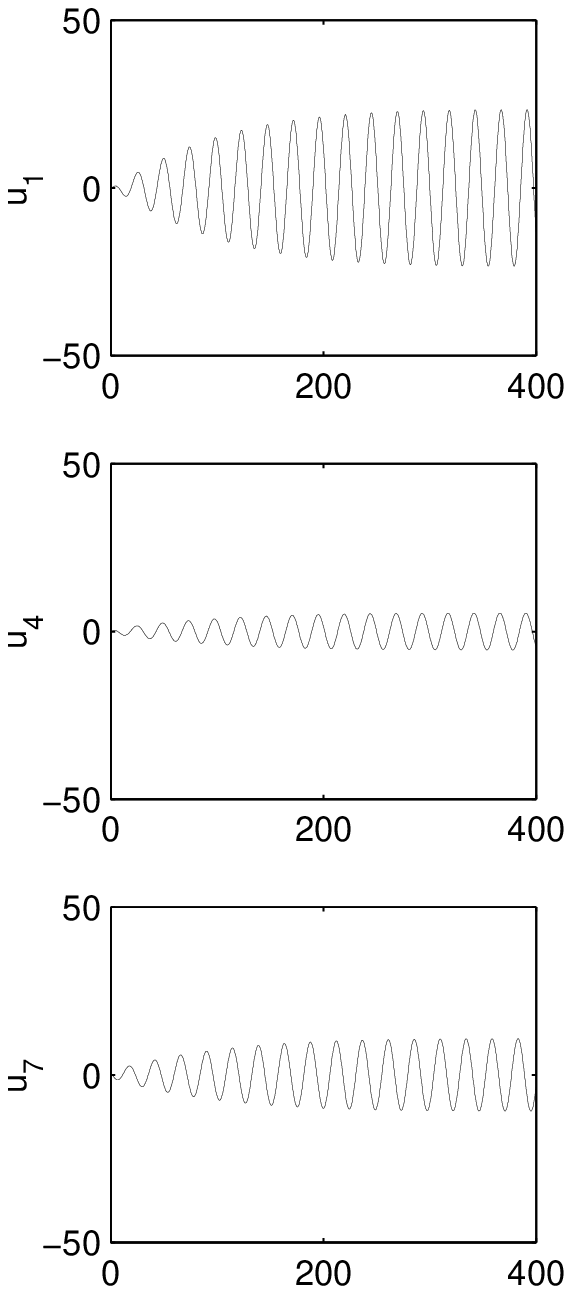}&
\epsfysize=6.6cm
\epsfbox{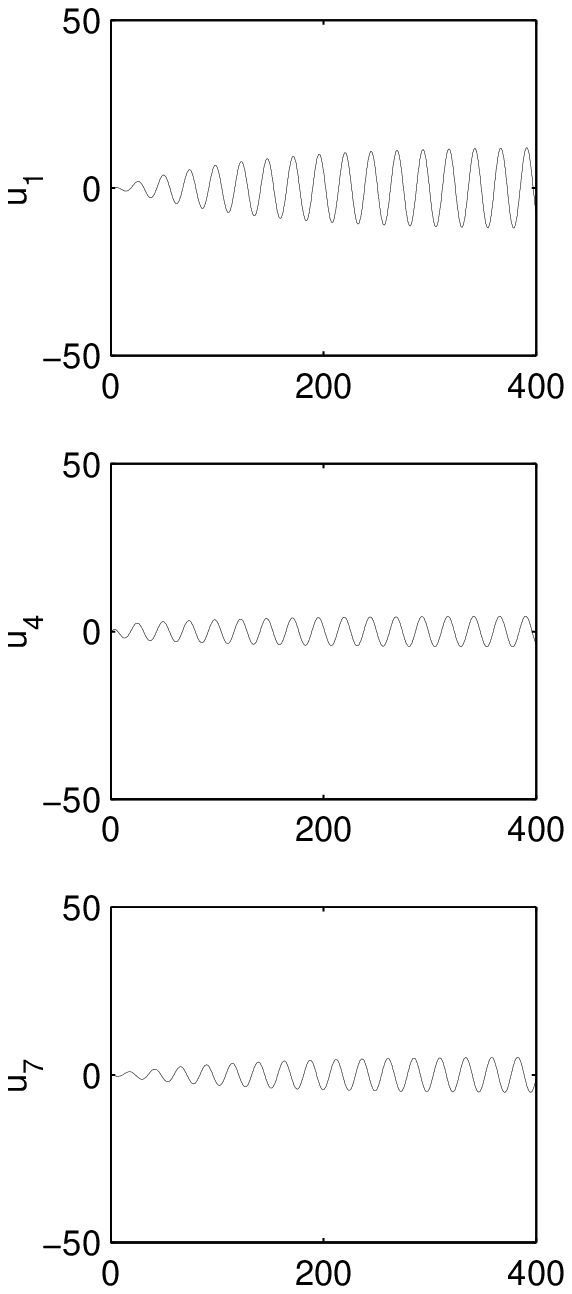}&  \\
\end{tabular}
\ec
\bc
\begin{tabular}{ccc}
  & B \hspace{2.0cm} & C \\
  & { \vspace{-0.3cm} \epsfysize=5.3cm  \epsfbox{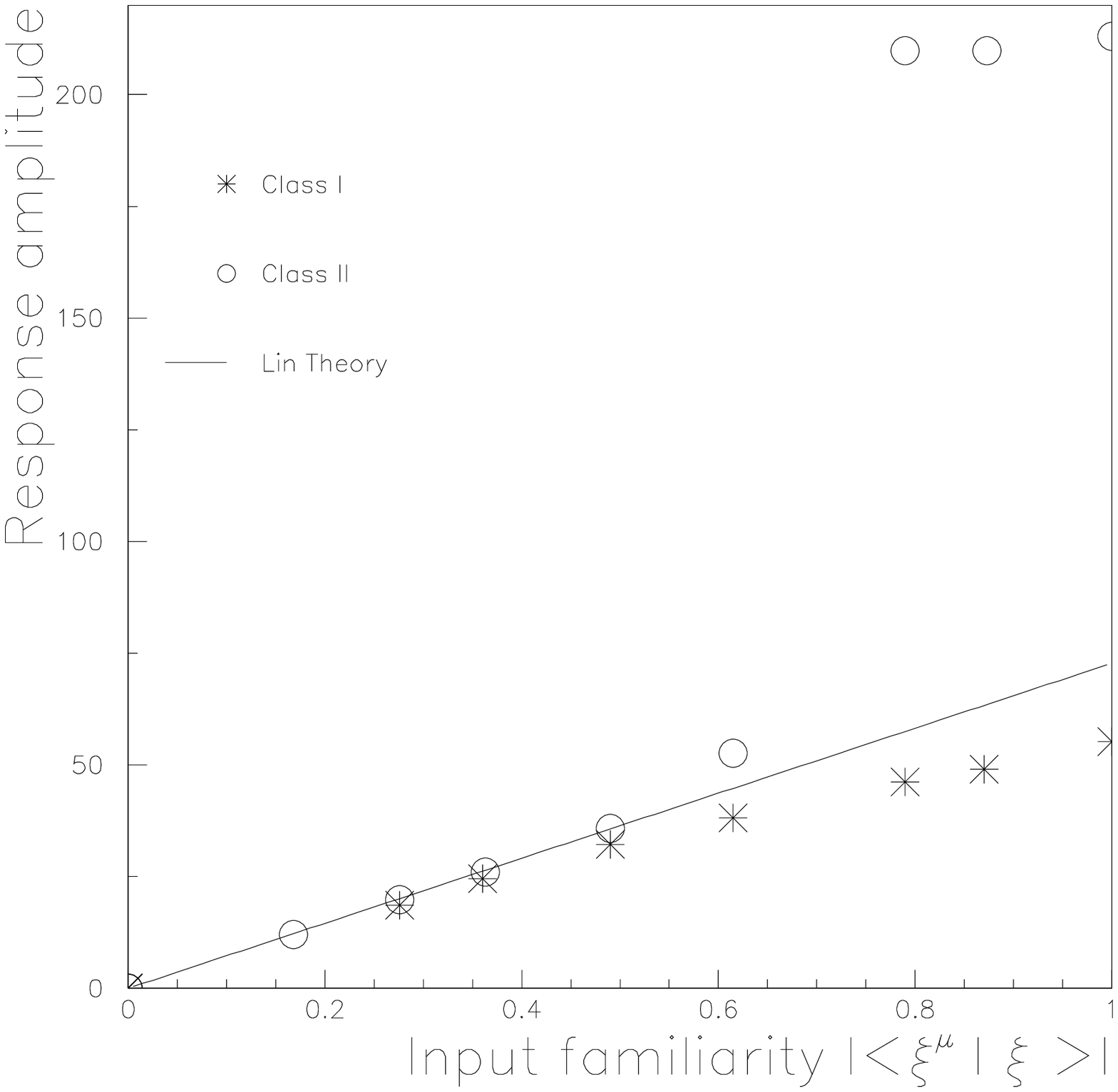}}\hspace{2.0cm} &  
 { \vspace{-0.3cm} \epsfysize=5.3cm  \epsfbox{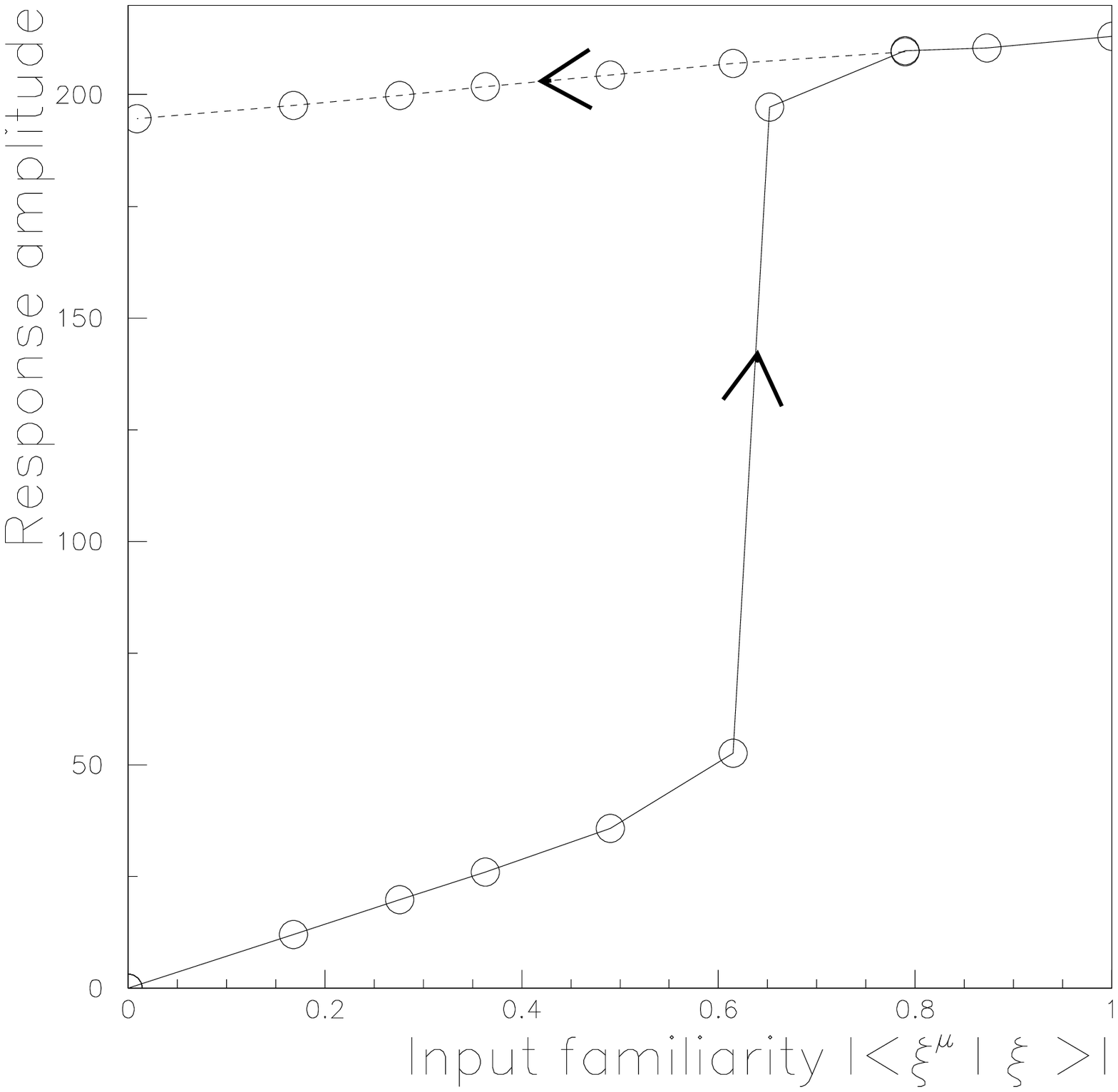} }
\end{tabular}
\vspace{0.1cm}    
\caption{
Pattern selectivity: \quad A: Response evoked on 3 of the 10 neurons of the network
by input patterns {\it a, b}, and {\it c} matching the imprinted
pattern {\it a} in frequency.
The class I activation functions shown in Fig.\ref{f_model}BD have been used.
\quad B: Response amplitude $|
\langle \bu ^+ |  \bxi^\mu \e^{-\i \omega_{\mu} t}
\rangle|$ vs. input overlap $|\langle \bxi {^{ {\mu}}}|
 \bxi \rangle|$, under input $ \delta \bI^+ = \bxi 
 \e^{-\i\omega_{ {\mu}} t} $.  
Results from
linearized theory (solid line) and from models with class I (stars) and II
(circles) nonlinearities.
\quad C: Hysteresis effects in class II
simulations. The response amplitude  depends on the history of the system: 
the output remains at the resonance level after input withdrawl 
(circles connected by dotted line).  
Circles connected by solid line correspond to the case of random or 
zero-overlap initial conditions.
The connecting lines are drawn for clarity only.
}
\label{fig5}
\ec
\end{figure}

We now consider an input $\delta \bI^+ {= \bxi \e^{-\i \omega t}}$ that
does not match the imprinted pattern $\bxi^\mu$.  In general, we can
decompose it into a component along $\bxi^\mu$ and a component in the
complementary subspace:
$\delta \bI^+ \equiv \delta \bI^+_\para + \delta \bI^+_\perp $, with 
$\delta \bI^+_{\para} \equiv  \langle \bxi^{\mu} | \bxi \rangle
\bxi^{\mu} \e^{-\i \omega t} \equiv N^{-1}(\sum_j \xi_j^{\mu *} \xi_j )
\bxi^{\mu} \e^{-\i \omega t}$.
then,
${\sf M }\bI^+ = \Pi(\omega;\omega_\mu) \bI^+_{\para}$,
and
\begin{equation}
\bu^+ = \chi (\omega; \omega_{\mu}) \delta \bI^+_{\para}
+ \chi_0 (\omega ) \delta \bI^+_{\perp}.       \label{decomp}
\end{equation}
The first term will be resonant at $\omega=\omega_\mu$, but the second
will not.  Thus, the system amplifies the component of the input along 
the stored pattern relative to the orthogonal one, as shown in Fig.~\ref{fig5}.  
Again, nonlinearity makes the response deviate from the linear prediction
at high response amplitudes, reducing and enhancing the responses for the
class I and II nonlinearities, respectively.  Class II nonlinearity also 
leads to hysteresis, with sustained responses even after the input is 
withdrawn, i.e., $|\langle {\bxi^{ {\mu}} }|\bxi\rangle |\rightarrow 0$.                     
The pattern selectivity can be measured by the ratio 
\begin{equation}
\frac{
|\chi (\omega_\mu ; \omega_{\mu})|}{|\chi_0 (\omega_\mu )|}
= 
\frac { \sqrt { (\alpha^2 + \beta \gamma - \omega_\mu^2)^2
                + ( 2\omega_\mu \alpha)^2 }}
{{\sqrt{ \Delta^2 + \epsilon^2}} }	   \label{selratio}
\end{equation}
where we used the resonance condition (\ref {res_c}).
When the input frequency $\omega$ deviates from $\omega_\mu$, the pattern 
selectivity ratio $|\chi (\omega ; \omega_{\mu})| / |\chi_0 (\omega)|$
is reduced.

\subsubsection{Interpolation and categorization.}

With multiple imprinted patterns, an input $\delta \bI^+ = 
\xi \e^{-\i \omega t}$ which overlaps with several of them
will evoke a correspondingly mixed resonant linear response $\bu^+ = 
\chi (\omega; \omega_{\mu}) \delta \bI^+_{\para}
+ \chi_0 (\omega ) \delta \bI^+_{\perp}$, where
$\delta \bI^+_\para = \sum_\mu \langle \bxi^{\mu} | \bxi \rangle
\bxi^{\mu} \e^{-\i \omega t}$.  That is,
any input in the pattern subspace produces a resonant linear response 
just like that to an input proportional to a single pattern.  This is a 
standard property of linear associative memories for orthogonal
patterns.  (We remind the reader that when the number of patterns is
much smaller than the number of units in the network, independent random
patterns may be taken as effectively orthogonal.)   This feature enables 
the system to interpolate between imprinted patterns, i.e., to perform
an elementary form of generalization from the learned set of patterns.
This property can be useful for input representation.  

\begin{figure}[ththththt!]
\setlength{\unitlength}{0.9pt}
\begin{picture}(400, 400) (0, 120)
\put(12, 553) {\bf A:}
\put(30, 370) {\epsfxsize=120pt \epsfbox{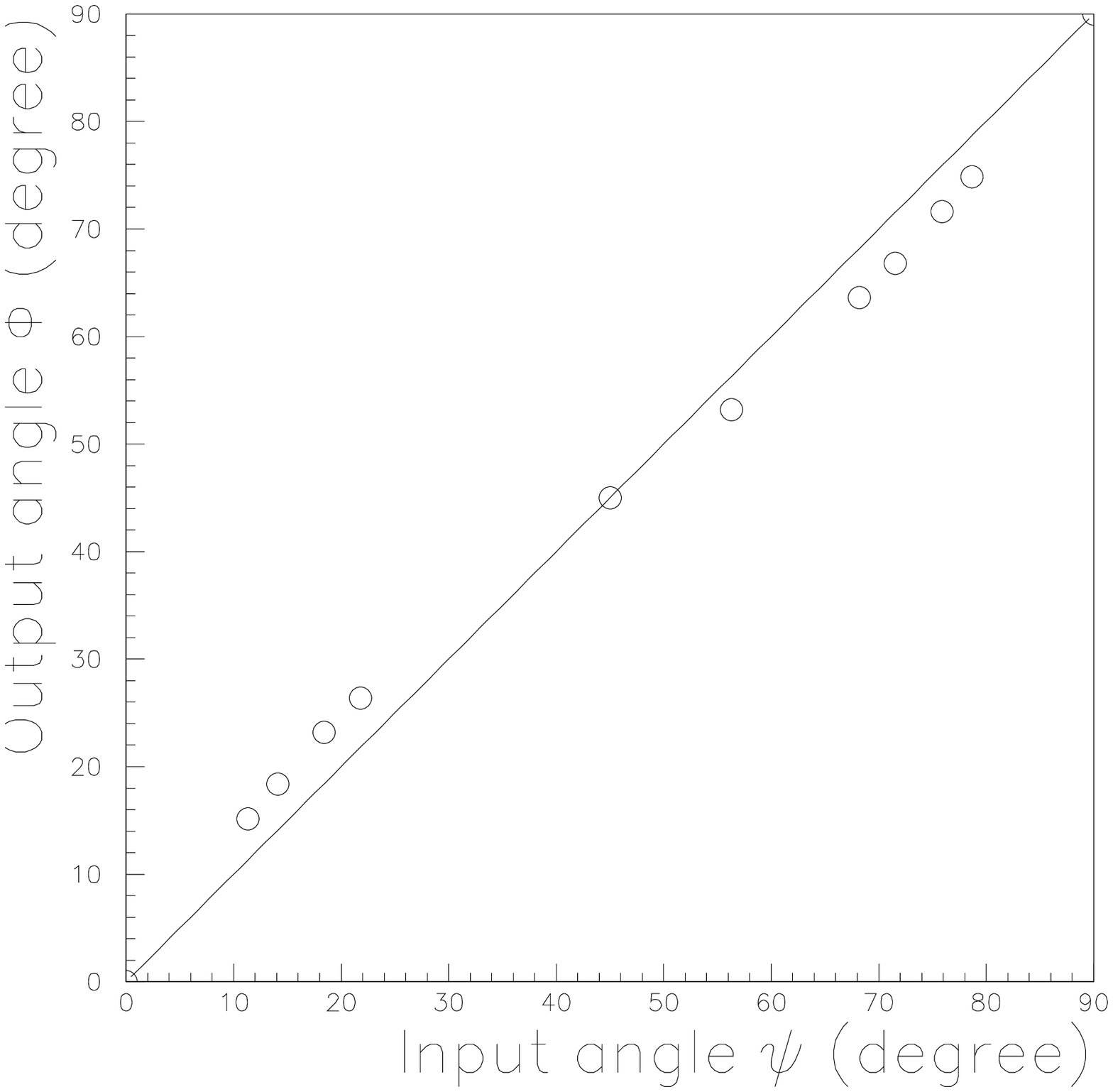}}
\put(212, 553) {\bf B:}
\put(230, 370) {\epsfxsize=120pt \epsfbox{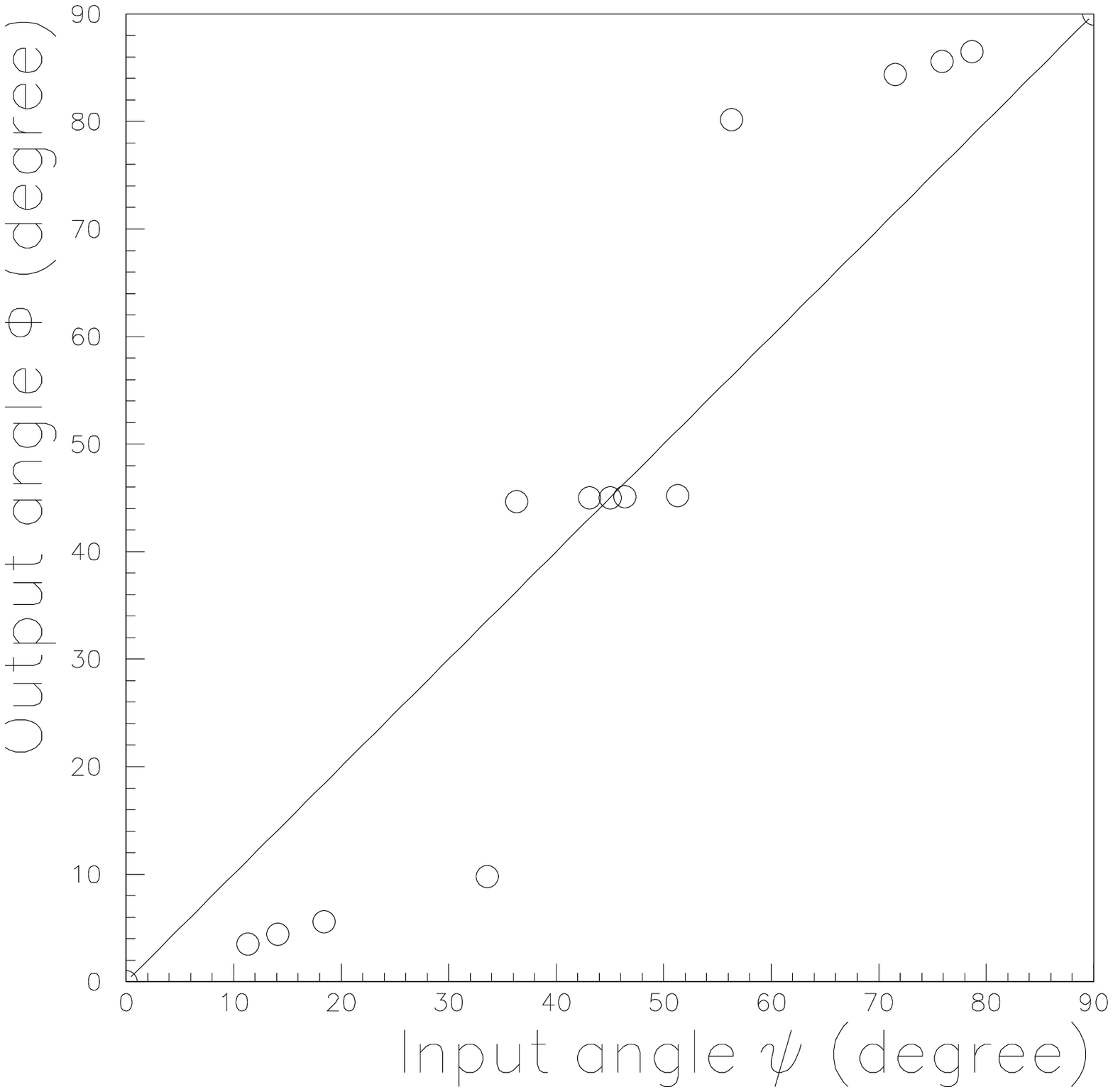}}

\put(12, 375) {\bf C:}

\put(21, 360) {$\omega = \omega_1$}
\put(21, 345) {$\bxi = \bxi^1$}
\put(-4, 120) {\epsfxsize=85pt \epsfbox{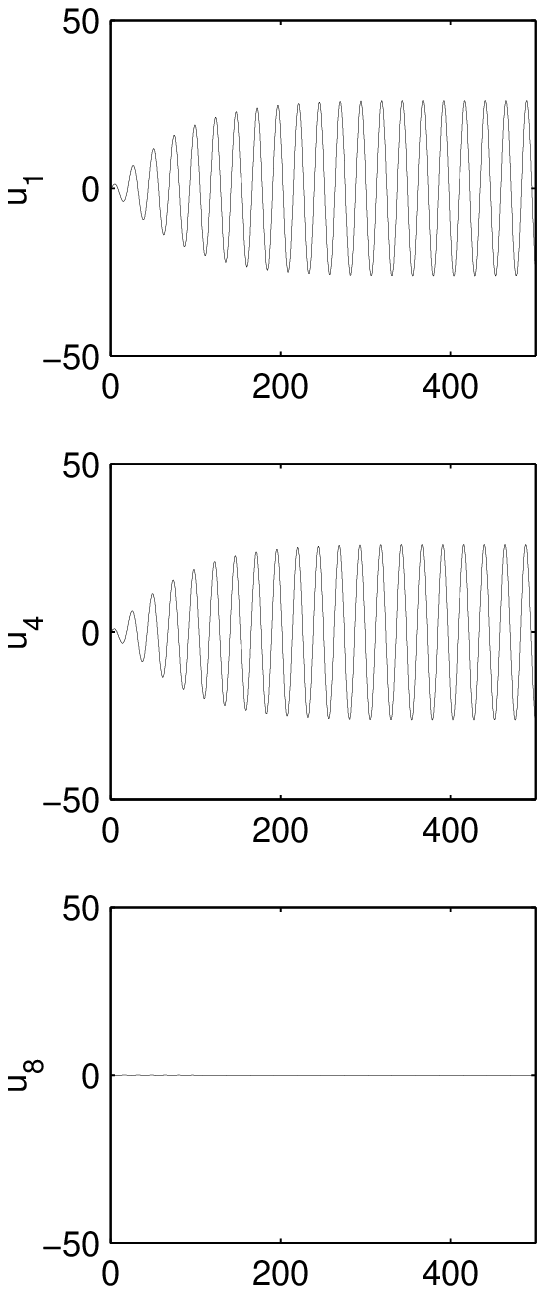}}

\put(119, 360) {$\omega = \omega_2$}
\put(119, 345) {$\bxi = \bxi^2$}
\put(94, 120) {\epsfxsize=85pt \epsfbox{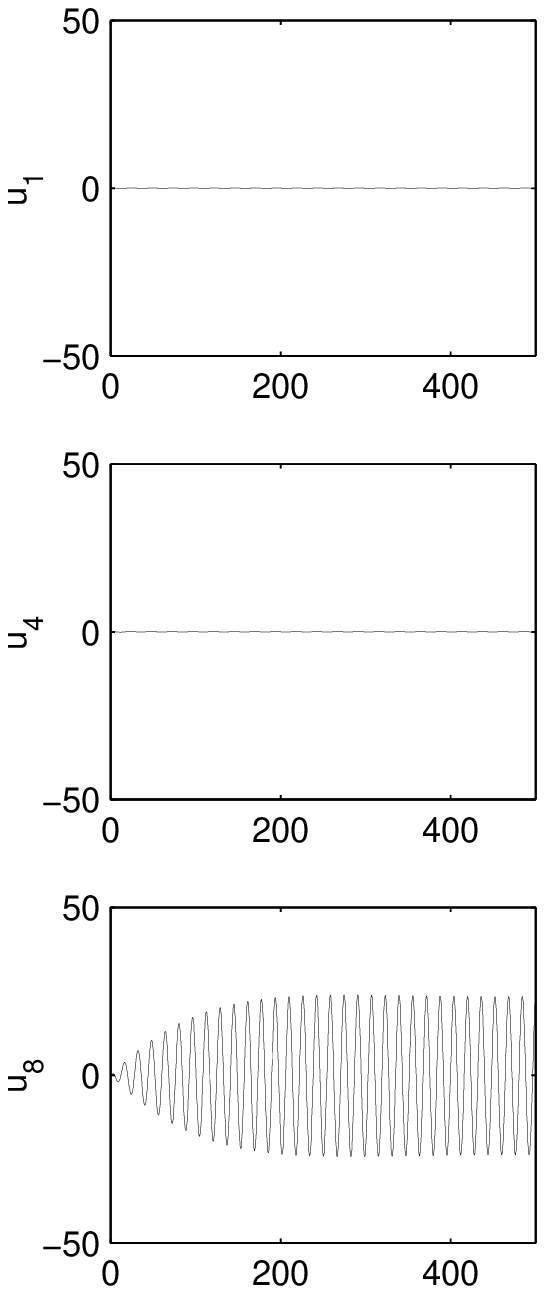}}

\put(217, 360) {$\omega = \omega_1$}
\put(217, 345) {$\bxi = \bxi^2$}
\put(192, 120) {\epsfxsize=85pt \epsfbox{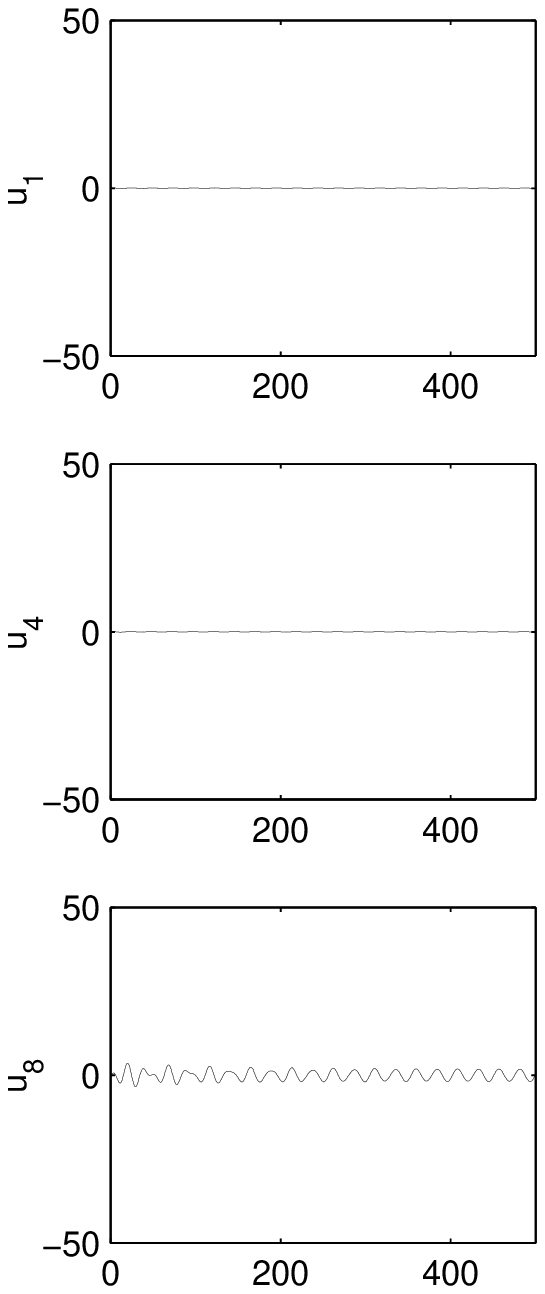}}

\put(315, 360) {$\omega = \omega_1$}
\put(315, 345) {$\bxi = a\bxi^1+b\bxi^2$}
\put(290, 120) {\epsfxsize=85pt \epsfbox{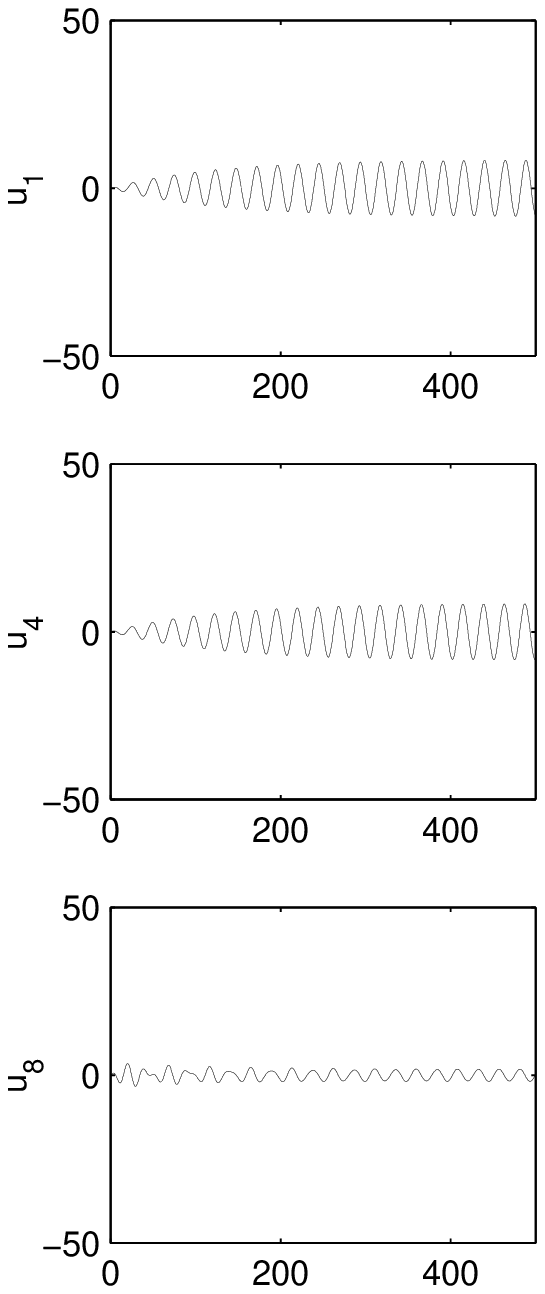}}
\end{picture}
\caption{
\label{fig6}
A, B: Input-output relationship when two
orthogonal patterns, $\bxi^1$ and $\bxi^2$, have been imprinted at the same
frequency  $\omega_\mu= 41 Hz$. 
Input 
$\bxi \propto \bxi^1\cos \psi  + \bxi^2\sin \psi$ and response
$\bu \propto \bxi^1\cos \phi + \bxi^2\sin \phi$.
Circles show
the simulation results, dotted lines show the analytical
prediction for the linearized model. A: Class I.
B: Class II.
C: Categorization using different imprinting frequencies. Plotted are  
responses of 3 of the 10 excitatory units to various input patterns and 
frequencies.
Patterns $\bxi^1 \e^{-\i \omega_1 t}$ and $\bxi^2 \e^{-\i \omega_2 t}$,
where
$\bxi^1 \perp \bxi^2$,  $\omega_1= 41$ Hz and $\omega_2 = 63$ Hz,
have been imprinted.
Matched kernels are used
with $\tilde A_{J} (\omega_1) =  0.5 - 0.025\i$ and
$\tilde A_{J} (\omega_2) =  0.5 - 0.43\i$ satisfying
resonance conditions. For the mixed input (fourth column),
$a= 1/\protect\sqrt{17}$ and $b= 4/\protect\sqrt{17}$.}
\end{figure}

A similar property also holds in the class I nonlinear model, but 
not in the class II model.  To see this, suppose the drive 
$\bI = \bxi e^{-\i \omega_\mu t} + \cc$ overlaps two imprinted 
patterns, with $\bxi \propto  \bxi^1 \cos \psi  + \bxi^2\sin \psi$, 
and write the response $\bu^+$ as $\bu^+ \propto \bxi^1\cos \phi  + 
\bxi^2\sin \phi$.  For a linear model, $\phi = \psi$. 
The class I nonlinear model
gives $45^o \ge \phi > \psi $ when $\psi < 45^o$ and 
$45^o< \phi < \psi $ when $\psi > 45^o$ (see Fig.~\ref {fig6}). Thus, it
tends to equalize the response amplitudes to $\bxi^1$ and 
$\bxi^2$ even when they contribute unequally to the input.   
In contrast, the class II nonlinear model amplifies the difference in
input strengths to give higher gain to the stronger input 
component, $\bxi^1$ or $\bxi^2$, thus performing a kind of categorization 
of the input.  Thus, the two nonlinearity classes leads to  different 
computational properties.
For the case shown in Fig.~\ref{fig6}B the parameters 
are such that the categorization is into three categories, corresponding
to outputs near $\xi^1$, $\xi^2$, and their symmetric combination.  For
stronger nonlinearity, bipartite classification is possible.

Another way to prevent undesirable interpolation between imprinted 
patterns (or classes of them) is to store different patterns or classes 
at frequencies that differ by more than the frequency tuning width.  
Suppose $\bxi^1 \e^{-\i\omega_1 t}$ and $\bxi^2 \e^{-\i\omega_2 t}$ 
are imprinted, with $\omega_1 \ne \omega_2$ and $\bxi^1 \cdot \bxi^2 
\approx 0$.  Then we have $J_{ij} = J_{ij}^1 + J_{ij}^2$ and
$W_{ij} = W_{ij}^1 + W_{ij}^2$, where $J_{ij}^\mu$ and $W_{ij}^\mu$ 
are given by equations (\ref {JWlearning}) with corresponding frequencies
for $\mu = 1, 2$.  The resonance and stability conditions should be enforced
separately for each pattern.

After learning, an input $\bI^+ = (\bxi^1 + \bxi^2) \e^{-\i \omega_1 t}$ 
at frequency $\omega_1$ will evoke a response
\begin{equation}
\bu^+ \approx \chi (\omega _1, \omega _1) \bxi^1
+ \chi (\omega_1, \omega_2) \bxi^2 \approx \chi (\omega _1, \omega _1) \bxi^1
\end{equation}
since $\chi (\omega_1, \omega_2) \ll \chi (\omega _1, \omega _1)$ by design
when $|\omega _1 - \omega_2| \ll \Delta$.  Hence, as illustrated in
Fig.~\ref{fig6}C,  the system filters out the oscillation patterns
learned at a different oscillation frequency from the input frequency.

\section{Nonlinear analysis}
\label{sec_nl}

Nonlinearity affects the response mainly at large amplitudes, which 
occur during resonant recall but not (we assume) in learning mode.  
Hence, in the following analysis, we leave the formulae for ${\sf J}$, 
${\sf W}$, and ${\sf M}$ unchanged, ignore nonlinearity in response
components orthogonal to the pattern subspace, and examine the corrections 
to the linear response $\bu = \chi (\omega ; \omega _\mu )\delta \bI_\para$.
We take the input to be along the imprinted pattern $\bxi^\mu$:
$\delta \bI^+_\para = I \bxi^\mu \e^{-\i \omega t}$. 
We focus on the nonlinearity in $g_u$, since
$g'_v$ only affects the local synaptic input, while $g'_u$
also affects the long-range input.
Equation (\ref {eqnsM}) then becomes  
\begin{equation}
[(\alpha - \i \omega )^2]{\bu}  =  
    {\M}  g_u (\bu ) 			 
 + (\alpha - \i \omega )\delta {\bf I}.                         
\label{eqnsMgu}
\end{equation}
where for simplicity, we include  $\gamma$ as a 
diagonal element of $\sf W$, and by $g_u(\bu)$ we 
mean a vector with components $[g_u(\bu)]_i = g_u(u_i)$. 
Making the ansatz
$\bu  = {q} \bxi^\mu \e^{-\i \omega t} 
+\cc +$ higher order harmonics,
we have,
\bea
u_j^3 &\approx& 
	 3 {q^2 q^*} {\xi^\mu_j}^2 \xi^{\mu *}_j \e^{-\i \omega t}  
  + \cc + \;\; \mbox{higher order harmonics},  
\eea
and analogously for $u_j^5$.  The quantity $q$ is the response amplitude 
of interest; in the linearized theory
$q\rightarrow \chi (\omega ; \omega _\mu ) I$.

For the two nonlinearity classes (\ref{gclass}) we have, respectively,
\bea
{\sf M}  \, g(\bu) &{\approx}& \left ( 
1 -  {3 a|{q}|^2} \sum_j {|\xi^\mu_j|}^4 
\right )
 {\sf M} \bu ,				 \label{Mg1} \\
{\sf M}  \, g(\bu) &{\approx }& \left ( 
1 + {3 a |{q}|^2} \sum_j {|\xi^\mu_j|}^4 -
{5 b |{q}|^4} \sum_j {|\xi^\mu_j|}^6 
\right ) {\sf M} \bu . 			\label{Mg2}
\eea
Thus, at a given response strength $|q|$, the imprinting strengths 
are effectively multiplied by the factors in parentheses.
Consequently, class I nonlinearity reduces the response at large 
amplitude, whereas class II nonlinearity enhances it as long as the
quadratic term in $|q|$ is larger than the quartic one. 

A consequence for class II is the fact that a system which is
very close to resonance ($\epsilon, \Delta \rightarrow 0$) in
the linear regime can become unstable at higher response levels.
The system will then jump to a new state in which the (negative) 
quartic term in (\ref{Mg2}) is large enough that stability
is restored, as seen in Figs.~\ref{freqtuning} and \ref{fig5}

Substituting (\ref{Mg1}) and (\ref{Mg2}) into {(\ref{eqnsMgu})} and matching
the coefficients of $\bxi^\mu \e^{-\i \omega t}$ on left and right sides, 
we obtain, for the two nonlinearity classes, respectively,
\begin{equation}
\chi^{-1} (\omega ; \omega _\mu ) q
+  {3 a B}  \sum_j |\xi^\mu_j|^4  |{q}|^2 {q}
= I,
\label{eqI}
\end{equation}
\begin{equation}
\chi^{-1} (\omega ; \omega _\mu )q - {3 a B} \sum_j |\xi^\mu_j|^4  |{q}|^2 {q}
+ { 5 b B}  \sum_j |\xi^\mu_j|^6  |{q} |^4 {q}
=  I.
\label{eqII}
\end{equation}
where   $B \equiv \Pi(\omega;\omega_\mu)/(\alpha - \i \omega ) $.
These equations can be solved for $q$.  It is apparent that in
general, both the phase and the amplitude of $q$ are modified by
the nonlinearity.

\section{Effects of synaptic weight constraints}
\label{gen_sec}
     
Because of the excitatory character of the pre-synaptic unit,
$J_{ij}$ and $W_{ij}$ connections have to be non-negative, a condition
not respected by our learning formula (\ref{JWlearning}) so far.
As a remedy, one may (1) add an initial background weight, $\bar J/N$ or 
$\bar W/N$, independent of $i$ and $j$, to each connection to make it 
positive, i.e.,
\begin{equation} 
\left\{
\begin{array}{l}
J_{ij} = \bar J/N + \sum _\mu 2\Re [\tilde A_{\rm J} \xi_i^\mu \xi_j^{\mu *}]/N
 \ge 0
							\\
W_{ij} = \bar W/N + \sum _\mu 2 \gamma \Re \left[ \frac{
\tilde A_{\rm W} \xi_i^\mu \xi_j^{\mu *}}{\alpha -\i \omega_\mu}\right]/N 
\ge 0 .							\\
\end{array} \label {addJWbar} \right . 
\end{equation} 
and/or (2) delete all net negative weights. 

It is clear from equations (\ref {addJWbar}) that adding a background weight 
is like learning an extra pattern $\bxi^0$ that is uniform and synchronous, 
with $\xi^0_i = 1$ for all $i$, with learning kernels 
$\tilde A^{(0)}_J(\omega_0)$ and $\tilde A^{(0)}_W(\omega_0)$ which satisfy
$2\Re \tilde A^{(0)}_J(\omega_0)$ = 1 and $2\gamma \Re [ A^{(0)}_W(\omega_0)
/\alpha-\i \omega_0] = 1$.   We assume that these kernels are the same as 
those with which the patterns $\xi_i^\mu$ are imprinted, up to an overall 
learning strength factor:  $\tilde A^{(0)}_{J,W}(\omega) = \eta \tilde A_{J,W}
(\omega)$, and that the imprinting frequencies are the same: 
$\omega_\mu = \omega_0$.  Thus if the $\xi_i$ are of unit magnitude, in order
to guarantee that no $J_{ij}$ or $W_{ij}$ are to be negative we need
$\eta \ge 1$.    

This strategy can be effective provided that the imprinting of the 
uniform extra pattern does not lead to violation of any stability condition.
Since we have assumed the imprinted patterns $\bxi^\mu$ ($\mu > 0$) are 
(roughly) orthogonal to $\bxi^0$, we can treat the extra pattern 
independently of the others, and we just have to satisfy the same stability 
conditions for it that we previously found for the imprinted patterns.  
That is, the singularities of $\chi(\omega;\omega_\mu)$ have to lie in the 
lower half of the $\omega$ plane, where now $\chi(\omega;\omega_\mu)$ 
(Eqn.~\ref{chi}) has to be computed from a $\Pi(\omega;\omega_\mu)$ which 
is a factor $\eta$ larger than before.  For $\eta \rightarrow 1$, 
we get no change in the stability conditions.

Nonnegativity can be more practically achieved by simply deleting 
the net negative weights.  For random patterns, and without background 
weights $\bar J$ and $\bar W$, this leads to deleting half of 
the weights $J_{ij}$ and $W_{ij}$ obtained from the learning rule, 
which weakens their effect, quantified by the function 
$\Pi(\omega;\omega_\mu)$, by a factor of 2.   In simulations we have
found that increasing the learning strength by this factor leads to
results like those found earlier when negative $J_{ij}$ and $W_{ij}$
were permitted. 

Finally, we remark that negative weights can also be simply implemented 
via inhibitory interneurons with very short membrane time constants.

\section{Summary and Discussion}

\subsection{Summary}

We have presented a model of learning and retrieval for associative 
memory or input representation in recurrent neural networks that exhibit 
input-driven oscillatory activities.  The model structure is an 
abstraction of the hippocampus or the olfactory cortex.
The learning rule is based on the synaptic plasticity observed
experimentally, in particular, long-term potentiation and long-term 
depression of the synaptic efficacies depending on the relative timing 
of the pre- and postsynaptic spikes during learning.  After learning, the
model's retrieval is characterized by its selective strong responses to 
inputs that resemble the learned patterns or their generalizations.
Our work generalizes the outer-product Hebbian learning 
rule in the Hopfield model to network states characterized by
complex state variables, representing both amplitudes and phases.
Our work differs from previous modeling in the following respects:
(1) We allow that stored patterns vary in both amplitudes and phases, 
as well oscillation frequency.
(2) We imprint input patterns into the synapses using a generalized 
Hebbian rule that gives LTP or LTD according to the relative 
timing of pre- and postsynaptic activity. 
(3) We explore two qualitatively different functions of the network: 
one (associative memory) is to classify inputs into distinct categories 
corresponding to the individual learned examples, and the other is to 
represent inputs as interpolations between or generalizations of learned 
examples.

The same model structure was used previously, with a conventional
Hebbian rule with $A_{J,W}(\tau)\propto \delta (\tau)$, by two of the 
authors in a model for odor recognition/classification and segmentation
in the olfactory cortex\cite{LH00}.  The principal new contributions in 
the current work are (1) linking the model with the recent experimental 
data on neural plasticity and LTP/LTD and dissecting the role of the 
functional form of the learning kernel $A_{J,W}(\tau)$ in determining 
the selectivity to input patterns and frequencies, (2) an extended 
analysis of input selectivity and tuning, (3) exploration of the two 
different computational functions (associative memory and input 
representation) of the model, and (4) a detailed analysis of nonlinearity 
in the model.

\subsection{Discussion}

By using both amplitude and phase to code information, it is possible 
either to encode additional information or to increase robustness
by redundantly coding the same information coded by the 
amplitudes.  Indeed, hippocampal place  cells, 
which code the spatial location of the animal, fire at different phases 
of the theta wave depending
on the location of the animal in the place fields\cite{Recce}.
In this case, the information encoding is redundant since the
location is in principle already encoded by the firing 
rates (i.e., oscillation amplitudes) in the neural population.
In our model, combined phase and amplitude coding requires
matching both the amplitude and phase patterns of the inputs with the
learned inputs under recall, making matching more specific.
This scheme necessitates learning both excitatory-to-excitatory
connections and excitatory-to-inhibitory ones.  Thus, in a system
of $N$ coupled oscillators, the stored items are coded by $2N$ variables ---
$N$ amplitudes and $N$ phases, requiring the specification of $2N^2$ 
synaptic strengths --- $N^2$ excitatory-to-excitatory synapses and another 
$N^2$ excitatory-to-inhibitory ones.  Omitting phase coding would require
learning of only $N^2$ synapses, e.g., of the excitatory-to-excitatory
connections, as in previous models\cite{Hendinetal,
WangBuhmannMalsburg}.

Our model's frequency selectivity adds additional matching specificity 
during recall. Furthermore, frequency matching can 
modulate the spiking timing reliability,
since higher or lower oscillation amplitudes, caused by
better or worse frequency matching, should make the firing
probabilities of the cells more or less modulated or locked by 
oscillation phases.
Frequency dependence of spike timing reliability 
has been observed in cortical pyramidal cells and 
interneurons~\cite{Sejnowski}.
In our model, the frequency tuning is a network property 
imprinted in long-range connections, although frequency tuning
as a resonance phenomenon could in principle exist in a single
neural oscillator or a local circuit.  

In our model, both excitatory-to-excitatory and 
excitatory-to-inhibitory syn\-apses are modifiable.  Experimentally,
there is as yet little evidence concerning plasticity 
in pyramidal-to-interneuron 
synapses.  More experimental investigations are needed.
In experiments by  Bell \cite{CurtisBell}, plasticity of the
excitatory-to-inhibitory synapses
between parallel fibers and medium ganglion cells in 
the cerebellum-like structure of the electric fish has been observed.
although these synapses are not part of a recurrent oscillatory circuit.

We explored the constraints on the learning kernel functions $A_{J,W}(\tau)$
imposed by the requirement of a resonant response.  A condition that
came up in almost all the variants of the model that we explored was
that $\tilde A_J'(\omega_\mu)$ should be positive in order to achieve
a strong, narrow resonance.  This means, roughly, that for 
excitatory-excitatory synapses LTP should dominate LTD in overall strength, 
for spike time differences smaller than $1/\omega_\mu$.

Another condition we considered was that the resonant frequency should 
be the same as the driving frequency $\omega_\mu$ during learning.  We saw 
that for real patterns and learning only of the excitatory-excitatory 
connections, this could not be satisfied for general $\omega_\mu$.  However,
with learning of the excitatory-to-inhibitory connections it could, for
a suitable (negative) value of $\tilde B''_W(\omega_\mu)$.  For complex
patterns (see Eqn.~\ref{epcomplexB}), the imaginary parts of both
$\tilde B_J$ and $\tilde B_W$ contribute to the shift, so if they have
opposite signs (of the correct relative magnitude) the condition can
be satisfied, independent of $\omega_\mu$.  These features should be
looked for in investigations of plasticity of excitatory-to-inhibitory
synapses. 

An interesting property we have identified in the model is its ability
to subserve two different computational functions: to {\it classify} inputs 
into distinct learned categories, and to {\it represent} input patterns as 
interpolation and generalizations of the prototype examples learned. 

Categorization is appropriate for associative memories and has been 
applied in our previous model of olfactory cortex\cite{LH00}.
In this context, interpolation between different learned patterns
is not desired; individually learned odors should have specific 
roles. It is more desirable to perceive individual odors within an 
odor mixture than to perceive an unspecific blend.
 
On the other hand, interpolation is advantageous in some circumstances.
Consider an animal learning an internal representation of a region of
space.  If particular spatial locations are represented as particular 
imprinted patterns, then locations in between them will be represented 
as linear combinations of these patterns.  Thus, the network is able
to represent a continuum of positions in a natural way.  Hippocampal 
place cells seem to employ such a representation.   
A network that interpolates can generalize from the
learned place fields to represent spatial locations between the
learned place fields by superposition of the neural activities
of the place cells.  Because the place fields are localized, the 
generalization is conservative (and thus robust):  It does not
extend beyond the spatial range of the learned locations or
to regions between distant, disjoint place fields.

We showed that our network can serve one or the other of these two
computational functions, depending on the nonlinearity in the neuronal
activation functions.  Class I leads to the interpolation or input 
representation operation mode, while class II leads to categorization.
The form of $g(u)$ could be subject to modulatory control, permitting
the network to switch function when appropriate.  The switch could
even be accomplished, for a suitable form of $g(u)$, simply by
a change in the DC input level, since it is possible to change the 
effective nature of the nonlinearity near the operating point by 
shifting the resting point.  It seems likely to us that the brain may 
employ different kinds and degrees of nonlinearity in different areas 
or at different times to enhance the versatility of its computations.

We have seen that it is possible to store different classes of patterns
at different oscillation frequencies, and that the network does not
interpolate between patterns stored at different frequencies.  This feature
gives the system the possibility of performing several different forms of 
input representation or categorization without interference between them.
For instance, all place fields could be stored at one frequency,
while odor memories could be stored at another, and there would be
no crosstalk between the two modalities if the frequencies differed
by much more than the resonance linewidth.  
 
In conclusion, we have seen that this rather simple network is endowed with
interesting computational properties which are consequences of the
combination of its oscillatory dynamics and the spike-timing-dependent
synaptic modification rule.  Although experiments to date have not clearly
uncovered examples of networks in the brain that function in just this 
fashion, we hope that our findings here will stimulate further investigations,
both theoretical and experimental.

\end{document}